# Strain effects on the binding and diffusion energies of Au adatoms and CeO$_2$ admolcules on Au, CeO$_2$, MgO and SrTiO$_3$ surfaces

Ahmad Ahmad, Ying-Cheng Chen, Jie Peng[‡], Anter El-Azab*

School of Materials Engineering, Purdue University, West Lafayette, IN 47907, USA

[‡]Was with Purdue University during this study

*Corresponding author: Anter El-Azab, aelazab@purdue.edu

**Abstract**

First-principles density functional theory (DFT) calculations were used to study the effects of elastic strains on the binding and diffusion activation energies of Au adatom and CeO$_2$ admolecule on Au (001), Ce-terminated CeO$_2$ (001), MgO (001), SrO- and TiO$_2$-terminationed SrTiO$_3$ (001) surfaces. In preparation for computing these energies, normal and shear strains within the range ±0.15% were applied in the plane of the surface of the supercell prior to placing the adsorbed species on the surface. Our study shows that the dependence of binding energies and diffusion barriers of adatoms and molecules on the strain varies significantly among surfaces. The strain was found to alter the symmetry of surface diffusion pathways causing anisotropy of the diffusion barriers. This strain-induced anisotropy depends on the orientation of the applied strains relative to the in-plane crystallographic directions of the free surface. The binding and diffusion activation energies were fit linearly in terms of strain components in the range ±0.15% and the extrapolated values compared favorably to DFT computed values up to ±0.5%. The scheme presented here for the computation and fitting of the binding and diffusion energies in terms of strain can be used to inform models of surface diffusion, clustering and growth of multi-component and multi-phase thin films and investigate the effect of strain on the self-organization in such systems.

## 1. Introduction

The impact of strain on the surface diffusion plays a significant role in self-organization of island during thin film growth. For example, the presence of misfit dislocations below the surface of a substrate, e.g., with a layered substrate, can influence the organization of islands due to coupling of diffusion with the resulting misfit strain at the top surface [1–4]. In other cases, the local strains arising due to the formation of heteroepitaxial islands was found to influence the self-organization of the islands themselves [5–7]. Strain also impacts the growth of vertically aligned nanocomposite (VAN) films [8–10]. Motivated by such observations, some authors used first-principles calculations to study the effect of strain on the activation barriers for exchange and hopping mechanisms of adatoms on surfaces [11–14]. Other authors exploited the dependence of the binding energies and diffusion barriers of adatoms on the state of the surface strains to explore preferential nucleation sites for island formation [12,13,15,16]. Elastic models of adatoms were also constructed. For example, similar to the elastic treatment of point defects, force dipole representations of adatoms were formulated, from which adatom coupling to the imposed strain can be computed [17,18]. To better understand island nucleation rates and pattern formation during thin film growth, it is important to understand how the surface strain influences binding and diffusion activation energies.

The current work is motivated by growth of thin-film configuration with vertically aligned phases (VAN systems). In such systems, the strain states can be complex since strain arises due to a multi-way lattice and thermal mismatch among all phases including the substrate [19,20]. In pillar-in-matrix VAN systems, both the in-plane and vertical mismatches and, hence, strain can be tuned by controlling the pillar volume fraction and density and selection of phase combinations [21,22]. The pillar morphology (shape and distribution) thus plays an important role [23]. Successful reports on the growth of oxide-metal [24], oxide-oxide [25], metallic alloy-oxide [26], and nitride-metal [27] as constituents of VAN systems were demonstrated using pulsed laser deposition (PLD). Evidence of direct impact of strains on the growth of VANs was shown in multilayer growth where growth of the metal pillars in the top layer is informed by the elastic strain field in the buffer layer due to the complex geometry and lattice mismatch in the bottom multiphase oxide-metal or oxide-oxide in the bottom layer [8–10].

At the mesoscopic level, using phase field model that incorporates elastic strain energy to describe multiphase film growth process under physical vapor deposition (PVD), it was shown that differences in the elastic stiffness of phases results in distinct microstructural morphologies [28,29]. The evolution of thin film morphology is controlled by adatoms diffusion and island growth kinetics. A report on the use of kinetic Monte Carlo (kMC) simulations to model heteroepitaxial growth in the early stages demonstrates that smaller (larger) misfit strain between the islands and the substrate leads to lower (higher) island density [30]. It is worth mentioning that increasing the misfit strain leads to high elastic energy at the island's periphery which increases the likelihood of adatoms detaching from islands relaxing the elastic strain [30]. In contrast, another study using kMC model that is coupled with the Green's function method to compute elastic strain energy in the islands due to the misfit suggests that including elastic strain effect decreases the binding energy of isolated adatoms [31]. This reduction leads to higher adatom diffusion, increasing the tendency of adatoms to cluster, which in turn increases (decreases) island size (density) [31].

It is crucial to quantify elastic interactions between adatoms and strained surfaces to understand growth mechanisms at the atomic scale. Hu and Ghoniem [14] studied the impact of biaxial surface stress of Tungsten adatom self-diffusion mechanisms, namely hopping, exchange and crowdion, on tungsten (001) and (110) surfaces using density functional theory (DFT). It was found that for hopping and exchange mechanisms, homogeneous tensile (compressive) load increases (decreases) the diffusion activation energy barrier of adatoms. Whereas, in the case of Crowdion diffusion, the opposite case is that tensile (compressive) load decreases (increases) the diffusion activation energy barrier of adatom. In the field of semiconductors and nanotechnology, van de Walle et al [11], using DFT, identified non-linear dependence of the diffusion barrier for Ge adatom hopping on Ge and Si (001) surfaces in which it increases (decreases) under imposed homogeneous negative (positive) biaxial strain. Another study reported that the dependence of adatom binding energies on surface strain can be either linear or non-linear which is subject to the bonding configuration of adatoms on the surface [32,33]. In an effort to highlight the drawbacks of polycrystalline substrate catalysts, an interesting approach that employs classical molecular mechanics simulations to study the binding energies of carbon adatom on various sites of Ni (111) that is characterized by inhomogeneous strained surface due to dislocations and grain boundaries shows that binding energies increase monotonically with increasing tensile strain [15].

In heteroepitaxial growth systems, including multiphase films, incorporating elastic strain effects into theoretical and numerical growth models is essential. KMC has been successfully applied to systems including oxide-metal [34,35], oxide-oxide [36–39], and multicomponent alloys [40,41]. In the model developed by Ahmad *et al* [35], despite its simplicity, it resulted in a more resolved representation of the pillars-in-matrix providing more detailed atomistic features of the complex pillar-matrix interface. This enables the calculation of elastic strains in the film accurately. The elastic strain field can then be used to interpolate the activation energy barrier of the diffusing adatoms within the kMC framework through constitutive relations describing elastic interactions between adatoms and elastic strain due to mismatch [42].

Motivated by experimental and modeling studies on Au(pillar)-$CeO_2$(matrix) film growth on $SrTiO_3$(substrate) [35,43] and multilayer oxide-metal pillar-matrix systems templated by $SrTiO_3$, MgO, and $CeO_2$ as buffer layers [9], the current work focuses on the diffusion of Au adatom and $CeO_2$ admolecule on various surfaces such as Au (001), $CeO_2$ (001), $SrTiO_3$ with SrO and $TiO_2$ (001) terminations, and MgO (001). The influence of surface elastic strain on the binding and diffusion of the adsorbed species is studied using DFT. In particular, we evaluate the effect of uniaxial and shear strains on the binding and diffusion activation energies of both Au adatom and $CeO_2$ admolecule across five different surfaces. Earlier studies [44–46] indicate that elastic strain can induce diffusion anisotropy in bulk systems. Keeping that in mind, we examine how strain influences diffusion barriers for multiple diffusion pathways along the [110], [100], [010] and [$\bar{1}$10] surface directions in the cubic systems we investigate.

This manuscript is organized as follows. A description of lattice strain and its impact on the binding and diffusion energies is given in Section 2. In Section 3, we explain the computational methodology used to perform DFT calculations. The results summarizing the influence of strain on the binding and diffusion energies of Au adatom and $CeO_2$ admolecule on different surfaces are presented and discussed in Section 4. Finally, Section 5 provides a brief summary of the current investigation and closing remarks on how the results can be used to parameterize thin film growth models.

## 2. Theoretical

### 2.1. Origin of elastic strain in multi-phase thin films

Let $A^0 = \{a_1^0, a_2^0, a_3^0\}$ denote a set of lattice translation vectors of unstrained crystal lattice. Under a homogeneous deformation, the deformation gradient $F_{ij}$ maps this reference lattice to its deformed state via [47]

$$A_{ij} = F_{iq} A_{qj}^0, \tag{1}$$

where the deformation gradient is defined by

$$F_{ij} = \delta_{ij} + \beta_{ij}, \tag{2}$$

with $\delta_{ij}$ being the Kronecker delta and $\beta_{ij}$ the lattice distortion, which is itself the displacement gradient, $u_{i,j}$. Inserting Eq. (2) into (1) and performing tensor inversion, the lattice distortion can be expressed in terms of the lattice parameters in the reference and distorted cases as follows

$$\beta_{ij} = [A_{il} - A_{il}^0] A_{lj}^{0-1}. \tag{3}$$

Here we consider single crystals subjected to infinitesimal strain states so that the rules of linear theory of elasticity apply. In this case, the strain is given by the symmetric part of the displacement gradient,

$$\varepsilon_{ij} = \text{Sym}(\beta_{ij}). \tag{4}$$

Following Gurtin's definition of surface strains [48,49], in the case of a flat surface with normal along, say, the $x_3$-direction, the strain components $\varepsilon_{11}, \varepsilon_{22}$ and $\varepsilon_{12}$ fall within the surface of interest. Fig 1(a) and (b), respectively, show the Au (001) surface under the uniaxial strain $\varepsilon_{11}$ and pure shear strain $\varepsilon_{12} = \gamma_{12}/2$, with $\gamma_{12}$ being the shear angle. The in-plane normal strain $\varepsilon_{11}$ and $\varepsilon_{22}$ in particular follow the formula:

$$\varepsilon_{ij} = \frac{\Delta a}{a^o} \delta_{ij}, \qquad (5)$$

with $\Delta a = (a - a^o)$ and $a^o$ and $a$ being the undeformed and deformed lattice parameter, respectively. A general strain state of the crystal can simultaneously result in non-trivial values of $\varepsilon_{11}$, $\varepsilon_{22}$ and $\varepsilon_{12}$. In the current work, it is assumed that the energy quantities of interest are mainly dependent on these three strain components. As will be shown later, the parameters of interest will be obtained by applying strains in the linear elastic regime. In this case, applying $\varepsilon_{11}$ ($\varepsilon_{22}$) to the relaxed lattice with lattice parameter $a^o$ results in deformed lattice parameter $a$ in the $x_1$ ($x_2$) directions, which is fed to the DFT calculations to preform analysis of structure and energies with the strained state of the crystal.

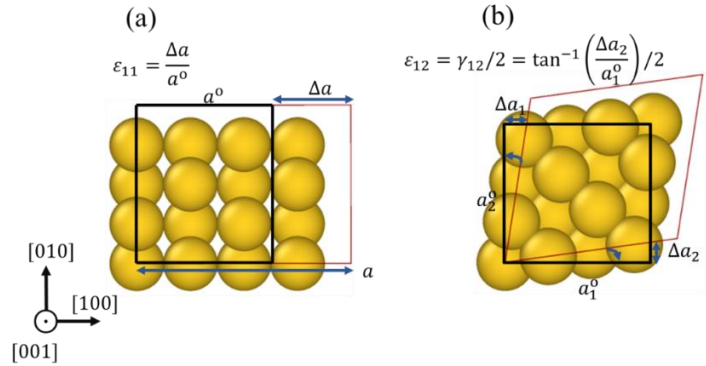

Fig. 1. Deformed Au lattice viewed down the [001] direction under (a) a uniaxial normal strain $\varepsilon_{11}$ and (b) shear strain $\varepsilon_{12} = \gamma_{12}/2$.

### 2.2. Elastic strain effects

In this section we develop expressions for the surface, binding and diffusion energies in terms of the strain in the linear elastic regime.

**Surface energies**

Consider a relaxed crystal volume $V$ that is subjected to an imposed homogeneous strain state $\varepsilon_{ij}$ at temperature $T$. The total free energy $\mathcal{F}[\varepsilon_{ij}, T]$ of the crystal can be expressed in the form

$$\mathcal{F}(\varepsilon_{ij}, T) = VF(0,T) + V\frac{\partial F(0,T)}{\partial \varepsilon_{ij}}\varepsilon_{ij} + \frac{V}{2}\frac{\partial^2 F(0,T)}{\partial \varepsilon_{ij}\partial \varepsilon_{kl}}\varepsilon_{ij}\varepsilon_{kl} \tag{6}$$

In the above, $F(0,T)$ is the free energy density at temperature $T$ and zero strain and the partial derivatives are evaluated at zero strain. At equilibrium, the first order derivatives $\frac{\partial F}{\partial \varepsilon_{ij}}$ vanish identically at zero strain since the stress is zero. The third term in Eq. (6) is elastic strain energy, with the second order derivatives $\frac{\partial^2 F}{\partial \varepsilon_{ij}\partial \varepsilon_{kl}}$ being the fourth rank elastic tensor $c_{ijkl}$ of the crystal. Expression (6) ignores terms of third order and higher. Imagine now that the relaxed crystal is cleaved to create a surface with normal $\boldsymbol{n}$ and area $A$. A surface energy term must now be added in the expression of the free energy of the crystal given in Eq. (6). Under a homogeneous strain state, the free energy expression can be rewritten in the form

$$\mathcal{F}(\varepsilon_{ij}, T; \boldsymbol{n}) = VF(0,T) + \frac{V}{2}\frac{\partial^2 F(0,T)}{\partial \varepsilon_{ij}\partial \varepsilon_{kl}}\varepsilon_{ij}\varepsilon_{kl} + A\gamma(0,\boldsymbol{n},T) + A\frac{\partial \gamma(0,T,\boldsymbol{n})}{\partial \varepsilon'_{\alpha\beta}(\boldsymbol{n})}\varepsilon'_{\alpha\beta}(\boldsymbol{n})$$
$$+ \frac{A}{2}\frac{\partial^2 \gamma(0,T,\boldsymbol{n})}{\partial \varepsilon'_{\alpha\beta}(\boldsymbol{n})\partial \varepsilon'_{\gamma\delta}(\boldsymbol{n})}\varepsilon'_{\alpha\beta}(\boldsymbol{n})\varepsilon'_{\gamma\delta}(\boldsymbol{n}), \tag{7}$$

where $\gamma(0,T,\boldsymbol{n})$ is surface energy density at zero strain. The fourth and fifth terms express the variation of the surface energy with the surface strain, $\varepsilon'_{\alpha\beta}$, and do not generally vanish in solids [50,51]. The fifth term in particular contains the surface elastic constants as the second order derivative of the surface energy with respect to surface strain [51]. The surface strain tensor $\varepsilon'_{\alpha\beta}$ (distinguished by the prime superscript and Greek indices) can be defined in terms of the bulk strains near a surface with normal $\boldsymbol{n}$ by applying the projection operator, $P_{ij}$, the surface strain is obtained [49]

$$\varepsilon'_{\alpha\beta} = P_{\alpha i}\varepsilon_{ij}P_{j\beta}, \tag{8}$$

with projection operator $P_{ij}$ defined by

$$P_{ij} = \delta_{ij} - n_i n_j. \tag{9}$$

In the current work, the normal $\bm{n}$ is always taken along the z-axis, i.e., $\bm{n} = \bm{e}_3$, which makes only the components $\varepsilon'_{11}, \varepsilon'_{12}$ and $\varepsilon'_{22}$ being the same as $\varepsilon_{11}, \varepsilon_{12}$ and $\varepsilon_{22}$, respectively. In the case of infinitesimal strain, the surface energy can be approximated in terms of strain up to first-order, thus the term containing the second order derivative is dropped. The first two terms in Eq. (7) can be written in the form: $\mathcal{F}(\varepsilon_{ij}, T) = VF(\varepsilon_{ij}, T)$. Rearranging Eq. (7) to isolate the surface energy,

$$\gamma\big(\varepsilon'_{\alpha\beta}(\bm{n}), T\big) = \frac{1}{A}\big(\mathcal{F}(\varepsilon_{ij}, T; \bm{n}) - VF(\varepsilon_{ij}, T)\big) = \gamma(0, T, \bm{n}) + \frac{\partial \gamma(0, T, \bm{n})}{\partial \varepsilon'_{\alpha\beta}(\bm{n})} \varepsilon'_{\alpha\beta}(\bm{n}). \tag{10}$$

In the above, $\frac{\partial \gamma}{\varepsilon'_{\alpha\beta}(\bm{n})}$ can be called the surface stress tensor.

**Binding energies of adatoms**

Consider an adatom or admolecule located at a local minimum energy site $\bm{r}^{\text{ad}}$ on the surface. The binding energy is defined as the energy required to separate the adatom from the free surface. The total free energy of the crystal with an adatom on its surface site at $\bm{r}^{\text{ad}}$ can be expressed as

$$\mathcal{F}(\varepsilon_{ij}, T; \bm{n}) = VF(0, T) + \frac{V}{2}\frac{\partial^2 F(0, T)}{\partial \varepsilon_{ij} \partial \varepsilon_{kl}} \varepsilon_{ij}\varepsilon_{kl} + A\gamma(0, \bm{n}, T)$$
$$+ A\frac{\partial \gamma(0, T, \bm{n})}{\partial \varepsilon'_{\alpha\beta}(\bm{n})} \varepsilon'_{\alpha\beta}(\bm{n}) + F^b\big(\bm{r}^{\text{ad}}, \varepsilon'_{\alpha\beta}(\bm{n}), T\big). \tag{11}$$

$F^b\big(\bm{r}^{\text{ad}}, \varepsilon'_{\alpha\beta}(\bm{n}), T\big)$ is the binding free energy of an adatom. The binding energy of adatom can be thus expressed by rearranging Eq. (11)

$$F^b\big(\bm{r}^{\text{ad}}, \varepsilon'_{\alpha\beta}(\bm{n}), T\big) = \mathcal{F}(\varepsilon_{ij}, T; \bm{n}) - A\gamma\big(\varepsilon'_{\alpha\beta}(\bm{n}), T\big) - VF(T, \varepsilon_{ij}). \tag{12}$$

Within the framework of linear elasticity, a crystal slab that is subjected to external deformation contributes an additional elastic interaction energy between the adatom and the strained surface. To first order approximation, and relative to the zero-strain state, the binding energy can be expressed as

$$F^b\left(r^{\text{ad}}, \varepsilon'_{\alpha\beta}(n), T\right) = F^b\left(r^{\text{ad}}, 0, T\right) + \frac{\partial F^b(r^{\text{ad}}, 0, T)}{\partial \varepsilon'_{\alpha\beta}} \varepsilon'_{\alpha\beta}. \tag{13}$$

Evaluated at zero strain, the term $\frac{\partial F^b(r^{\text{ad}},0,T)}{\partial \varepsilon'_{\alpha\beta}}$ is the intrinsic surface stress induced by the adatom, which is equivalent to a dipole tensor based on the theory of elasticity [18,46,52,53]. By writing $F^b\left(r^{\text{ad}}, \varepsilon'_{\alpha\beta}(n), T\right) = E^b\left(r^{\text{ad}}, \varepsilon'_{\alpha\beta}(n), T\right) - TS^b\left(r^{\text{ad}}, \varepsilon'_{\alpha\beta}(n), T\right)$, and ignoring the entropic part, the current work will focus on evaluating the binding internal energy $E^b$ only at zero temperature. This quantity will be evaluated by the same expression (13) in which $F^b$ is replaced with $E^b$.

**Atomic jumps**

The diffusion mechanism considered in this work is the hopping mechanism in which an adatom or admolecule jumps between two low energy surface sites in a given direction, $d$. According to the transition state theory, the hopping rate along $d$ is given by the expression [54,55]

$$\Gamma^d = \frac{k_B T}{h} \frac{Z^{\text{ts}}}{Z^{\text{gs}}}, \tag{14}$$

where $k_B$ is Boltzmann constant, $h$ Planck's constant, and $Z^{\text{ts}}$ and $Z^{\text{gs}}$ are the partition functions at the transition and ground (initial) states of direction $d$, respectively [56–58]. The partition function itself is expressed in the form

$$Z = \exp -\frac{\mathcal{F}}{k_B T}. \tag{15}$$

Note that the Helmholtz free-energy is given by $\mathcal{F}(\varepsilon_{ij},T) = \mathcal{E}(\varepsilon_{ij},T) - T\mathcal{S}(\varepsilon_{ij},T)$, where $\mathcal{E}$ is the total internal energy, including the elastic strain energy contribution, and $\mathcal{S}$ is the total entropy. With this in mind, inserting (15) into (14) and utilizing the definition of the free energy density, the hopping rate can be rewritten as,

$$\Gamma^d = \Gamma^o e^{-\frac{(\mathcal{E}^{ts}-\mathcal{E}^{gs})^d}{k_B T}} = \Gamma^o e^{-\frac{\Delta E^d}{k_B T}}. \tag{16}$$

In the above, $\mathcal{E}^{ts}$ and $\mathcal{E}^{gs}$ are the internal energies of the surface/adatom (admolecule) system at the transition and ground states, respectively, for the direction $d$, with $\Delta E^d = \mathcal{E}^{ts} - \mathcal{E}^{gs}$ being activation energy, and $\Gamma^o$ is the attempt frequency. The latter is $\Gamma^o = k_B T/h\, e^{\Delta S^d/k_B}$ with $\Delta S^d = \mathcal{S}^{ts} - \mathcal{S}^{gs}$ being the hopping entropy. Formally speaking, a complete determination of the hopping rate requires the determination of $\Delta E^d$ and $\Delta S^d$. Here, we assume $\Gamma^o$ to be a numerical factor on the order of $10^{12}$ to $10^{13}$ Hz, and thus only focus on the activation barrier, $\Delta E^d$.

Considering a homogeneous strain state at the surface and the expression of the diffusion activation energy, $\Delta E^d(\varepsilon_{ij},T) = \mathcal{E}^{ts}(\varepsilon_{ij},T) - \mathcal{E}^{gs}(\varepsilon_{ij},T)$. The transition and ground state energies can be expressed in the form: $\mathcal{E}^{ts}(\varepsilon_{ij},T) = \mathcal{E}^{ts}(0,T) + \frac{\partial \mathcal{E}^{ts}(0,T)}{\partial \varepsilon'_{\alpha\beta}} \varepsilon'_{\alpha\beta}$ and $\mathcal{E}^{gs}(\varepsilon_{ij},T) = \mathcal{E}^{gs}(0,T) + \frac{\partial \mathcal{E}^{gs}(0,T)}{\partial \varepsilon'_{\alpha\beta}} \varepsilon'_{\alpha\beta}$. With this in mind, $\Delta E^d(\varepsilon_{ij},T)$ can be expressed in the form

$$\begin{aligned}\Delta E^d(\varepsilon_{ij},T) &= \mathcal{E}^{ts}(\varepsilon_{ij},T) - \mathcal{E}^{gs}(\varepsilon_{ij},T) \\ &\approx \Delta E^d(0,T) + \left[\frac{\partial \mathcal{E}^{ts}(0,T)}{\partial \varepsilon'_{\alpha\beta}} - \frac{\partial \mathcal{E}^{gs}(0,T)}{\partial \varepsilon'_{\alpha\beta}}\right] \varepsilon'_{\alpha\beta},\end{aligned} \tag{17}$$

where $\Delta E^d(0,T) = \mathcal{E}^{ts}(0,T) - \mathcal{E}^{fs}(0,T)$ and $\left(\frac{\partial \mathcal{E}^{ts}(0,T)}{\partial \varepsilon'_{\alpha\beta}}\right)$ and $\left(\frac{\partial \mathcal{E}^{ts}(0,T)}{\partial \varepsilon'_{\alpha\beta}}\right)$ are evaluated at zero strain. These two derivatives are the adatom surface stress which depends on the configuration of adsorbed species and the surface. The last formula can be specialized to the case of $T = 0$, which is computed with DFT in the current work.

## 3. Computational methodology

All DFT calculations were performed using the Vienna Ab initio Simulation Package (VASP) [59]. The exchange-correlation functional of Perdew-Burke-Ernzerhof (PBE) [60] has been used under the generalized gradient approximation (GGA) [61]. The Kohn-Sham equations [62] were solved by treating the electron-ion interactions within the projected augmented wave (PAW) approach [63]. The GGA method was used for all materials, Au, MgO, and SrTiO$_3$ except for systems involving CeO$_2$ where GGA+$U$ was employed, with $U = 3$ eV, as suggested by other studies [35,64,65]. For both bulk and slab models, the cutoff energy value for the plane-wave basis set used for the Au metal is 250 eV, whereas for oxides, CeO$_2$, MgO, and SrTiO$_3$, a value of 450 eV was used. The Monkhorst-Pack [66] k-point mesh for sampling the Brillouin zone (BZ) used in the calculation is 8×8×8 and 4×4×1 for bulk and slab systems, respectively. For determining the partial occupancies of orbitals, the Methfessel-Paxton smearing widths [67] for the Au metal system are set to 0.2 eV whereas a value of 0.05 eV was set for the Gaussian smearing scheme for the oxide systems, CeO$_2$, MgO and SrTiO$_3$. The stopping criteria for the convergence of the self-consistent electronic loop and ionic relaxation in the bulk models (prior to constructing slabs with free surfaces) are chosen to be $1 \times 10^{-6}$ eV and $1 \times 10^{-4}$ eV/Å, respectively.

After obtaining fully relaxed bulk systems, 2×2 (001)-plane slabs are created by adding a vacuum with a thickness of 15 Å in the [001] direction to eliminate the interactions between the free surface and the back side of the image slab. Stoichiometry was maintained in constructing oxide slabs. The CeO$_2$ surface considered in this work exhibits Ce termination only, whereas the SrTiO$_3$ surface exhibits both SrO and TiO$_2$ terminations. With periodic boundary conditions applied parallel to the surface along [100] and [010] directions, and while fixing the lower two layers while allowing upper layers to relax in all degrees of freedom, the convergence criteria are set to such that the total energy is less than $1 \times 10^{-4}$ eV while the average force acting on every atom is $1 \times 10^{-2}$ eV/Å.

As shown in Fig. 2, on the Au (001), $CeO_2$ (001) and MgO (001), SrO (001) and $TiO_2$ (001) terminated surfaces, the Au adatom and $CeO_2$ admolecule are placed at the equilibrium sites, referred to as hollow sites. Since $CeO_2$ admolecule is assumed to behave as a rigid molecule during diffusion, it is reasonable to assume that the 4-folded-symmetric hollow site represents the equilibrium position for both the Au adatom and $CeO_2$ admolecule. After obtaining all relaxed configurations, the calculations of the diffusion activation energies of the hopping mechanisms along direction $d$ are performed using the nudged elastic band (NEB), ensuring a complete repeating unit, as implemented in VASP [68,69]. The NEB calculations were done via constructing 5-7 intermediate images and using spring force of -5 eV/Å. For FCC Au, $CeO_2$, and MgO surfaces, we consider hopping along the four directions [100], [010], [110], and [$\bar{1}$10]. For diffusion on $SrTiO_3$ with both SrO and $TiO_2$ terminations, the directions [100], [010], and [110] were considered. As a simplification, the $CeO_2$ admolecule was assume to maintain a configuration along the migration path that is the same as that at the initial and final sites.

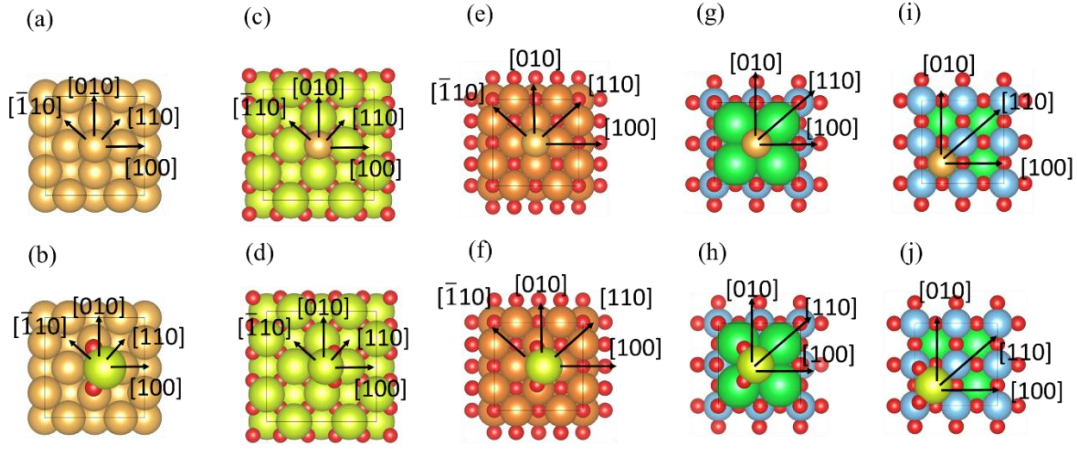

Fig. 2. Initial configurations of Au adatom and CeO$_2$ admolecule for 4 directional hopping diffusion mechanisms, on (a,b) Au (001), (c,d) Ce-terminated CeO$_2$ (001), (e,f) MgO (001), (g,h) SrO- and (i,j) TiO$_2$-terminated SrTiO$_3$ (001) surfaces. Crystallographic axes are oriented with [001] out of plane, [100] to the right and [010] points in the upward direction.

Uniaxial and shear strains in the range -0.15% to 0.15% were imposed on the relaxed bulk and slab supercells. For the strained slab, we assume that fixing the bottom two layers and relaxing the top layers allows for mimicking mixed boundary conditions on the supercell where the free surface becomes traction-free and the bottom layer conforms to the bulk away from the surface [16,51]. In the sequel, and for simplicity, we do not distinguish between $\varepsilon_{ij}$ and $\varepsilon'_{ij}$. Ignoring the entropic part of the surface free energy $\gamma(\varepsilon_{ij})$, the internal energy part $E^s(\varepsilon_{ij})$ can be expressed in the form

$$E^s(\varepsilon_{ij}, \boldsymbol{n}) = \frac{1}{2A}\left(E^{\text{slab}}[\varepsilon_{ij}] - \frac{N^{\text{slab}}}{N^{\text{bulk}}}E^{\text{bulk}}[\varepsilon_{ij}]\right), \tag{18}$$

with $T$ assumed to be 0 K. In the above, we adopted the DFT notation for surface energy. The factor of $1/2$ corresponds to the creation of 2 surfaces in the slab model. $E^{\text{slab}}[\varepsilon_{ij}]$ and $E^{\text{bulk}}[\varepsilon_{ij}]$ are the energies of the slab and bulk supercells, respectively, and $N^{\text{slab}}$ and $N^{\text{bulk}}$ are the total number of atoms in the slab and bulk, respectively, which may or may not be the equal. Similarly, the binding energy at a specific surface lattice site can be obtained using Eq. (12) and ignoring the entropic part,

$$E^b(\mathbf{r}^{\mathrm{ad}}, \varepsilon_{ij}) = E^{\mathrm{tot}}[\varepsilon_{ij}] - E^{\mathrm{slab}}[\varepsilon_{ij}] - E^{\mathrm{ad}}, \tag{19}$$

where $E^{\mathrm{tot}}[\varepsilon_{ij}]$ is the energy of the slab and adatom and $E^{\mathrm{ad}}$ is the energy of the adsorbed Au atom or CeO$_2$ molecule. After obtaining the binding energy at every strain state, a best linear fitting is obtained to construct the adatom surface stress (strain derivative term) in Eq. (13). We remark here that the FCC lattice symmetry must be considered when constructing $\frac{\partial E^b}{\partial \varepsilon_{ij}}$, that is $\frac{\partial E^b}{\partial \varepsilon_{11}} = \frac{\partial E^b}{\partial \varepsilon_{22}}$ and $\frac{\partial E^b}{\partial \varepsilon_{12}} = \frac{\partial E^b}{\partial \varepsilon_{21}}$ in the case of placing an adatom on a hollow site. For the CeO$_2$ molecule, this may not be true, but we ignore the differences. The activation barriers $\Delta E^d(\varepsilon_{ij})$ can be expressed as

$$\Delta E^d[\varepsilon_{ij}] \approx \Delta E^d[0] + \frac{\partial \Delta E^d}{\partial \varepsilon_{ij}} \varepsilon_{ij}. \tag{20}$$

Again, exploiting symmetry, the components we are after are $\frac{\partial \Delta E^d}{\partial \varepsilon_{11}}, \frac{\partial \Delta E^d}{\partial \varepsilon_{22}}$ and $\frac{\partial \Delta E^d}{\partial \varepsilon_{12}}$ for all hopping directions. Note that applying uniaxial strain along [100] results in anisotropy in the diffusion barrier along [100] and [010] yielding to $\frac{\partial \Delta E^{[100]}}{\partial \varepsilon_{11}}$ not being equal to $\frac{\partial \Delta E^{[010]}}{\partial \varepsilon_{11}}$. On the other hand, the diffusion activation energies along [110] and [$\bar{1}$10] directions satisfy $\frac{\partial \Delta E^{[110]}}{\partial \varepsilon_{11}} = \frac{\partial \Delta E^{[\bar{1}10]}}{\partial \varepsilon_{11}}$. Applying shear strain results in a derivative $\frac{\partial \Delta E^{[110]}}{\partial \varepsilon_{12}}$ not equal to $\frac{\partial \Delta E^{[\bar{1}10]}}{\partial \varepsilon_{12}}$. Thus, generally speaking, care must be taken as to how the surface directional symmetry is or is not helping reduce the number of quantities to be computed using NEB method when strain is applied.

## 4. Results and discussion

### 4.1. Surface energies

Performed as explained above, surface energy calculations were validated by comparing the relaxed lattice constants and surface energies at zero strain with published values. As Table 1 shows, the results are in good agreement with values reported in the literature. For SrTiO$_3$ surface with TiO$_2$ termination, we notice that the surface energy slightly deviates from literature values, likely due to methodological differences, particularly the use of PBE functional in this work.

Table 1. Calculated lattice parameters and surface energies of the materials considered in the current work.

| Material | Method | Bulk lattice constant (Å) | Surface | Method | $E^s$ (J/m$^2$) |
|---|---|---|---|---|---|
| Au | This work | 4.156 | Au (001) | This work | 0.877 |
| | LDA | 4.090 [70] | | LDA | 1.343 [71] |
| | PBE | 4.170 [70] | | PBE | 0.864 [72] |
| CeO$_2$ | This work | 5.452 | Ce (001) | This work | 4.181 |
| | PW91 | 5.423 [73] | | MD | 4.200 [74] |
| | PAW | 5.390 [75] | | | |
| MgO | This work | 4.246 | MgO (001) | This work | 0.923 |
| | LDA | 4.167 [76] | | Exp | 0.970 [77] |
| | PBE | 4.273 [76] | | PBE | 0.860 [78] |
| SrTiO$_3$ | This work | 3.948 | SrO (001) | This work | 1.073 |
| | LDA | 3.860 [79] | | B3PW | 1.150 [80] |
| | PBE | 3.940 [79] | | | |
| | P3PW | 3.904 [79] | TiO$_2$ (001) | This work | 1.077 |
| | B3PW | 3.910 [79] | | B3PW | 1.290 [80] |

Fig. 3 shows the variation of surface energy $E^s$ with strains $\varepsilon_{11}$ and $\varepsilon_{12}$ for various materials. For the Au surface, the increase of $E^s$ with $\varepsilon_{11}$ from compression to tension is consistent with the results shown by Elsner *et al* [51]. It is believed that this trend is attributed to the residual surface stress which results in atoms at the surface being in a compressed state. As such, imposing external compression provides additional work in favor of the surface stress whereas tensile strain adds work against it leading to increase in the surface energy. The linear interpolation seems consistent with Eq. (10) because at small strains, it is permissible to neglect higher order terms as the linear term dominates. As shown in Fig. 3(a), the surface energies of Au, MgO, SrTiO$_3$ with SrO- and TiO$_2$ terminations all exhibit a rising linear trend, unlike Ce-terminated CeO$_2$ surface that exhibits a linear trend with negative slope. This suggests that CeO$_2$ surface possess tensile residual surface distortion in contrast to the compressive residual surface distortions of Au, MgO and SrTiO$_3$ surfaces. While the effect of uniaxial strain on surface energy is easily detectable, applying shear strain in the regime of linear elasticity produces only minor changes in surface energy, at the 4$^{th}$ or 5$^{th}$ decimal place. As shown in Fig. 3(b), all energies are symmetric with respect to shear strain, which can be attributed to the cubic symmetry of the surfaces. Interestingly, for MgO, SrO, and TiO$_2$ surfaces exhibit local minima and maxima points with pure shear imposed.

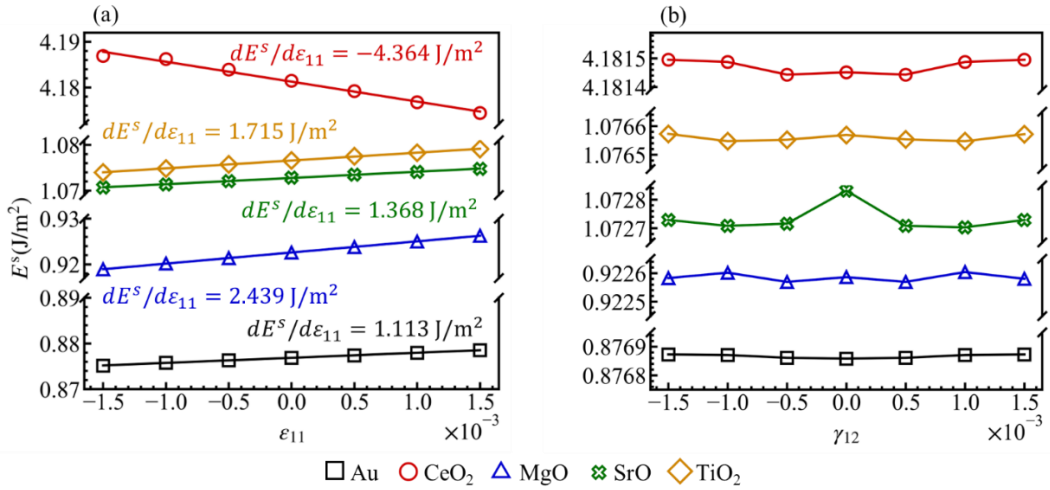

Fig. 3. Surface energies as a function of (a) uniaxial and (b) pure shear strain for Au (001), Ce- terminated CeO$_2$ (001), MgO (001), SrO- and TiO$_2$-terminated SrTiO$_3$ (001) surfaces.

From the linear interpolation (solid lines) shown in Fig. 3(a), we extract $\frac{\partial E^s}{\partial \varepsilon_{ij}}$. By symmetry, $\frac{\partial E^s}{\partial \varepsilon_{11}} = \frac{\partial E^s}{\partial \varepsilon_{22}}$ and because the shear contribution is negligible, it is assumed that $\frac{\partial E^s}{\partial \varepsilon_{12}} = \frac{\partial E^s}{\partial \varepsilon_{21}} = 0$. The surface energy can thus be expressed in the form

$$E^s(\varepsilon_{11}, \varepsilon_{22}) = E^s(0) + \frac{\partial E^s}{\partial \varepsilon_{11}} \varepsilon_{11} + \frac{\partial E^s}{\partial \varepsilon_{22}} \varepsilon_{22} . \qquad (21)$$

The DFT data of surface energy as a function of strains in the range ±0.15% were used to fit the linear dependence of surface energy on strain. Comparing the fit values to the DFT calculations at larger strains, ±0.5%, confirms that the linear dependence works up to at least the latter. The results are summarized in Tables S1-S4. It is expected though that higher order terms might be necessary to include in Eq. (21) at larger strains.

**4.2. Binding energies**

Figs. 4 and 5 show the relaxed structures with the Au adatom and $CeO_2$ admolecule located at hollow sites, respectively. The corresponding binding energies at zero strain are listed in Table 2, and they are in good agreement with reported values in the literature. Notably, the differences in $E^b$ of Au and $CeO_2$ on $SrTiO_3$ whether $TiO_2$- or SrO-terminated surfaces arise from the choice of adsorption sites. Looking at Fig. 5, it is interesting to note that $CeO_2$ admolecule induces larger displacement in the surface atoms than the Au adatom does.

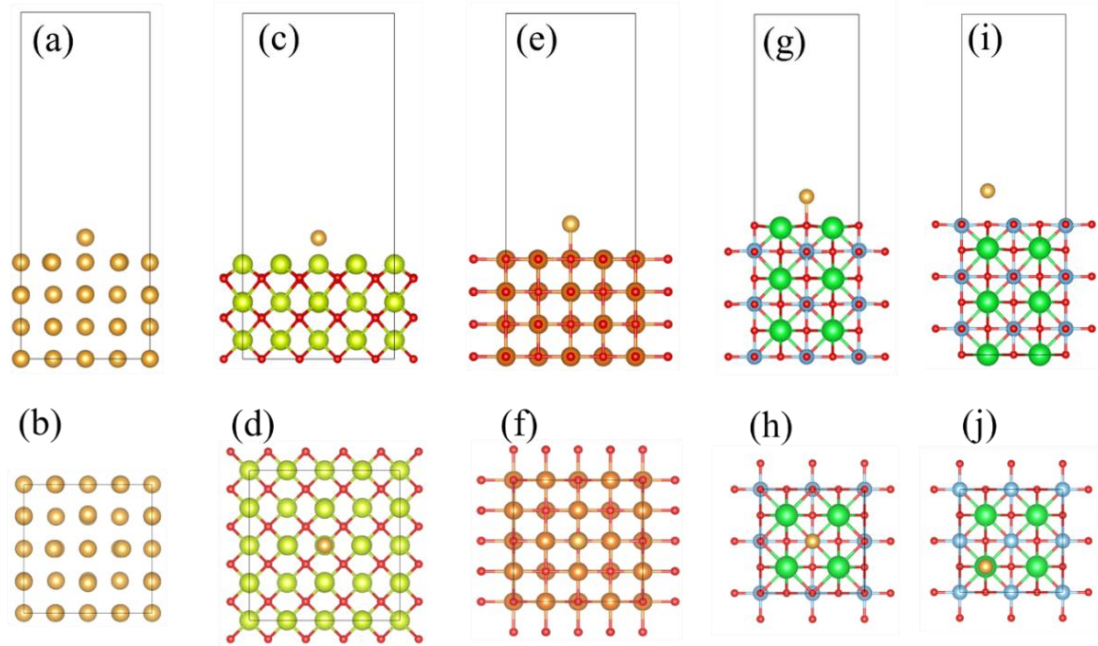

Fig. 4. Relaxed configurations of Au adatom on (a,b) Au (001), (c,d) Ce-terminated $CeO_2$ (001), (e,f) MgO (001), (g,h) SrO- and (i,j) $TiO_2$-terminated $SrTiO_3$ (001) surfaces. The top and lower rows, respectively, show vertical cross-sectional and top views with adatom positioned at hollow sites. In the top row, [100] points to the right, [010] points into the page and [001] points upward. In the lower row, [100] points to the right, [010] points upwards and [001] direction points out of the page.

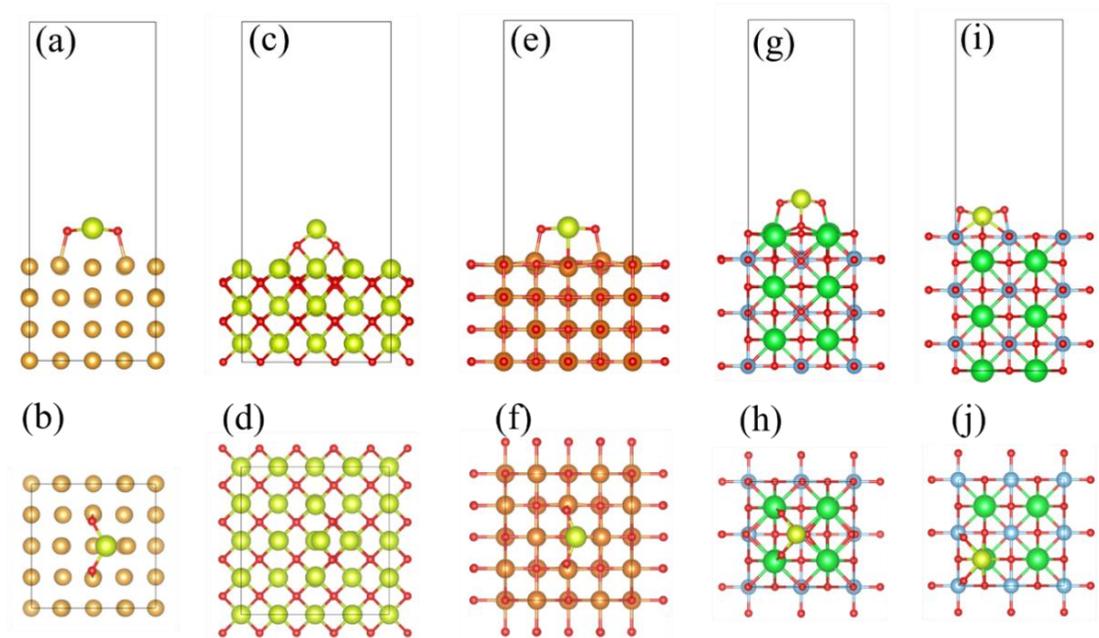

Fig. 5. Relaxed configurations with CeO$_2$ admolecule placed on (a) Au (001), (c) Ce-terminated CeO$_2$ (001), (e) MgO (001), (g) SrO- and (i) TiO$_2$-terminated SrTiO$_3$ (001) surfaces. Top view showing the relaxed configurations with CeO$_2$ admolecule placed at the hollow sites on (b) Au (001), (d) Ce-terminated CeO$_2$ (001), (f) MgO (001), (h) SrO- and (j) TiO$_2$-terminated SrTiO$_3$ (001) surfaces. In the top row, [100] points to the right, [010] points into the page and [001] points upward. In the lower row, [100] points to the right, [010] points upwards and [001] direction points out of the page.

Table 2. Binding energies of Au adatom and CeO$_2$ admolecule at the minimum energy site on the surfaces of interest and in the absence of applied strain.

| Surface | Adsorbed specie | Method | $E^b$ (eV) |
|---|---|---|---|
| Au (001) | Au | This work | -2.981 |
| | | PBE [35] | -2.994 |
| | | EAM [81] | -3.45 |
| | CeO$_2$ | This work | -1.267 |
| | | PBE [35] | -1.234 |
| CeO$_2$ (001) | Au | This work | -4.183 |
| | | PBE [35] | -4.335 |
| | CeO$_2$ | This work | -4.504 |
| | | PBE [35] | -4.565 |
| MgO (001) | Au | This work | -1.005 |
| | | PW-91 [82] | -0.890 |
| | | PBE [83] | -1.090 |
| | CeO$_2$ | This work | -3.420 |
| SrO (001) | Au | This work | -1.369 |
| | | PBE [35] | -1.521 |
| | | PBE [84] | -1.370 |
| | CeO$_2$ | This work | -4.431 |
| TiO$_2$ (001) | Au | This work | -0.462 |
| | | PBE [35] | -0.903 |
| | | PBE [84] | -0.600 |
| | CeO$_2$ | This work | -6.383 |
| | | PBE [35] | -0.370 |

Fig. 6 shows the binding energy as a function of strain evaluated using Eq. (19) for Au adatom and CeO$_2$ admolecule, respectively. Linear fitting can be constructed, with the slope representing adatom or admolecule-induced surface stress, $\frac{\partial E^b}{\partial \varepsilon_{ij}}$, in the limit of small strain. Because of lattice symmetry, we assume that $\frac{\partial E^b}{\partial \varepsilon_{11}} = \frac{\partial E^b}{\partial \varepsilon_{22}}$. Similar to what is observed on the effect of the shear strain on surface energies in Fig. 3(b), there does not seem to be a well-defined linear fit of the binding energies as function shear strain. In general, because the binding energy change with shear strain is a lot smaller than the uniaxial, we set $\frac{\partial E^b}{\partial \varepsilon_{12}} = \frac{\partial E^b}{\partial \varepsilon_{21}} \approx 0$.

In Fig. 6(a), it is observed that Au adatom favors tensile strain at the Au surface, whereas on the Ce-terminated $CeO_2$ surface, compressive strains are energetically more favorable. (The words favor and favorable are used to refer to stronger bonding). Likewise, from Fig. 6(c), we notice that $CeO_2$ admolecule favors the tensile strain at the Au surface and compressive strains at the Ce-terminated $CeO_2$ surface. It is also noticed that the binding of Au adatom and $CeO_2$ admolecule on $SrTiO_3$ with SrO-terminated surface behave differently; Au adatoms favor compressive strains whereas $CeO_2$ admolecules favor tensile strain. On MgO and $TiO_2$-terminated $SrTiO_3$ surface, both Au adatom and $CeO_2$ admolecule favor tensile strain. In Fig. 6(b) and (d), the impact of shear strain on the binding of Au and $CeO_2$ species is negligible. (The observed changes are in the order of $10^{-6}$ to $10^{-5}$ eV.) As the elastic strain impacts the energetics of bonding of adatoms and admolecules, film growth process can be made more favored or less favored by tuning the applied strain. The distinct response of adsorbed species to surface strain thus suggests that the local strain arising from lattice mismatch in multiphase systems may drive phase separation during thin film growth [9,43,47,85].

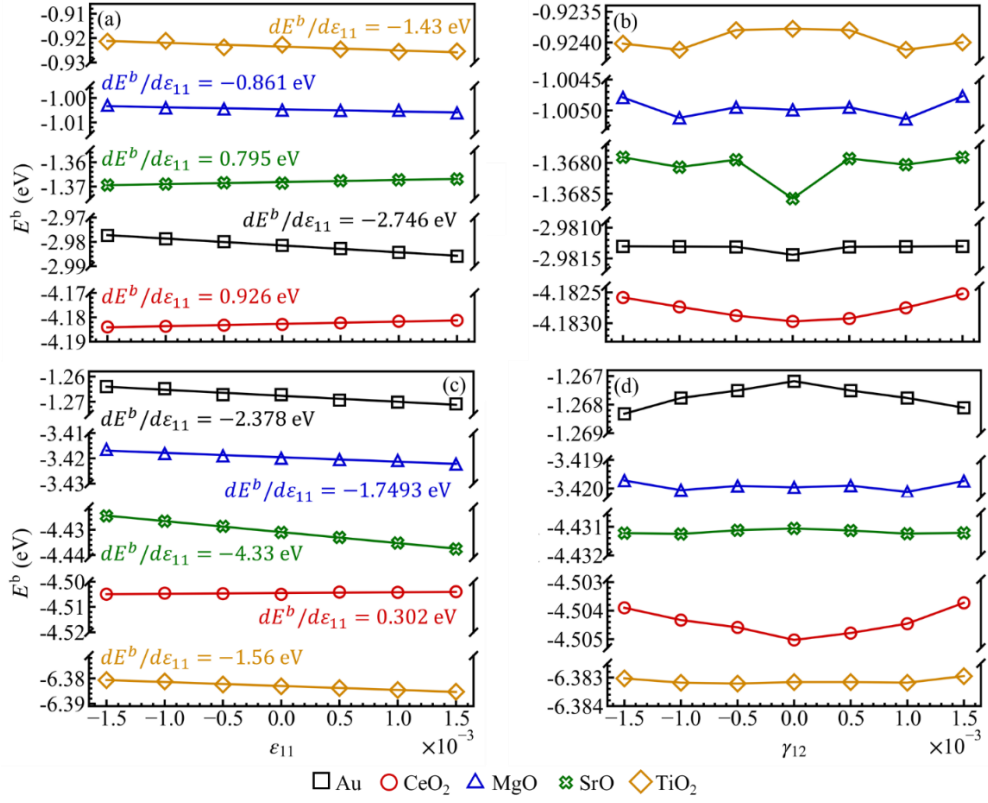

Fig. 6. Binding energies of Au adatom (a,b) and $CeO_2$ admolecule (c,d) at hollow sites on various surfaces as functions of strain. Panels (a,c) and (b,d) correspond to uniaxial and pure shear strain, respectively. The symbols refer to the surface material.

After dropping terms related to the shear strain, the linear fitting of the binding energy in terms of strain is expressed in the form:

$$E^b(\varepsilon_{11}, \varepsilon_{22}) = E^b(0) + \frac{\partial E^b}{\partial \varepsilon_{11}}\varepsilon_{11} + \frac{\partial E^b}{\partial \varepsilon_{11}}\varepsilon_{22}. \tag{22}$$

The values of $\frac{\partial E^b}{\partial \varepsilon_{11}}$ for every system is shown as the slope of best linear fit shown in Fig. 6(a) and (c). The fitting was performed using data obtained by applying strains in the range -0.15% to 0.15%. The linear fit was then extrapolated to strain values of 0.5% and -0.5%, and the results reasonably agree to DFT values of the binding energies at these relatively large strains; see Table S5-S8 for further details. In the case of $CeO_2$ admolecule on $SrTiO_3$ with both terminations, the extrapolated values and DFT evaluated values at the large strain values differ in first or second decimals, implying that the second order corrections of the form $\frac{\partial^2 E^b}{\partial \varepsilon_{11} \partial \varepsilon_{11}}$, $\frac{\partial^2 E^b}{\partial \varepsilon_{22} \partial \varepsilon_{22}}$ and $\frac{\partial^2 E^b}{\partial \varepsilon_{11} \partial \varepsilon_{22}}$ might be required for better fitting. Such refinement is, however, beyond the scope of this work.

### 4.3. Activation barriers for diffusion

The activation barriers, $\Delta E^d$, of Au adatom and $CeO_2$ admolecule at relaxed state on five surfaces and three directions, [110], [100], and [010], are reported in Table 3. These diffusion activation energies are in good agreement with values reported in the literature up to $\pm 0.1$ eV. For the surfaces of Au, $CeO_2$ and MgO surfaces, diffusion along [110] direction has a lower activation barrier than [100] and [010]. For SrTiO3 with SrO termination, Sr atoms form a square lattice on the surface, with oxygen atoms in the middle. As such, when the adatom or admolecule is placed on top of the oxygen atom, as shown in Fig. 5(g) and (h), the activation barriers along [100] and [010] are lower than those along [110]. We remark here that NEB calculations of $CeO_2$ admolecule hopping along [110] on $TiO_2$-terminated SrTiO3 surface failed due to instability of the intermediate configuration in which the $CeO_2$ admolecule sits on top of the Ti atom. We attribute this to the fact that the $CeO_2$ molecule is assumed to be rigid. We thus inform the reader that the impact of strain on $CeO_2$ diffusion along [110] on $TiO_2$-terminated SrTiO3 surface is not considered in this study for this reason. Having said so, the assumption that the $CeO_2$ molecule is rigid can be relaxed in subsequent investigations.

Table 3. Activation barriers for Au adatom and CeO$_2$ admolecule hopping on different surfaces in the absence of external strain.

| Surface | Adatom | Diffusion direction | $\Delta E^d$ (eV) | $\Delta E^d$ (eV)[Ref.] | Admolecule | Diffusion direction | $\Delta E^d$ (eV) | $\Delta E^d$ (eV) [Ref.] |
|---|---|---|---|---|---|---|---|---|
| Au (001) | Au | [110] | 0.5668 | 0.64 [81] | CeO$_2$ | [110] | 0.3306 | |
| | | [100] | 1.1422 | 1.24 [35] | | [100] | 0.6700 | 0.77 [35] |
| | | [010] | 1.1422 | | | [010] | 0.6916 | |
| CeO$_2$ (001) | Au | [110] | 0.9554 | 0.98 [35] | CeO$_2$ | [110] | 1.2458 | |
| | | [100] | 1.1141 | 1.06 [35] | | [100] | 1.5701 | 1.58 [35] |
| | | [010] | 1.1142 | | | [010] | 1.5101 | |
| MgO (001) | Au | [110] | 0.1667 | 0.24 [86] | CeO$_2$ | [110] | 1.1043 | |
| | | [100] | 0.3687 | 0.47 [86] | | [100] | 1.4657 | |
| | | [010] | 0.3687 | | | [010] | 1.4654 | |
| SrO (001) | Au | [110] | 0.7577 | 0.75 [35] | CeO$_2$ | [110] | 2.8436 | |
| | | [100] | 0.5365 | 0.57 [35] | | [100] | 1.5363 | |
| | | [010] | 0.5371 | | | [010] | 1.5363 | |
| TiO$_2$ (001) | Au | [110] | 0.1675 | 0.26 [35] | CeO$_2$ | [100] | 1.0022 | 1.02 [35] |
| | | [100] | 0.2688 | 0.30 [35] | | [010] | 1.0511 | |
| | | [010] | 0.2701 | 0.33 [35] | | | | |

Fig. 7 shows the minimum energy paths along the hopping direction of both Au (left column) adatom and CeO$_2$ admolecule (right column) on Au (001) surface as computed using the NEB method. For hopping along [110], Fig. 7(a) and (b), only a slight variation in the activation barrier is observed when uniaxial strain is applied in the [100] direction. Under the same strain condition, the energy profile for Au adatom transition in along the [100] direction becomes sensitive to strain as shown in Fig. 7(c). Moreover, it can be easily observed that the energy profiles of transitions along the [010] direction start to develop local energy minimum on opposite sides of the peak as seen in Fig. 7(e). CeO$_2$ admolecule diffusion along [100] or [010], however, shows negligible changes in the energy landscape in the hopping direction compared to the unstrained case. Finally, under applied shear, the energy profiles for both Au adatom and CeO$_2$ admolecule along [110] are essentially indistinguishable, Fig. 7(g) and (h).

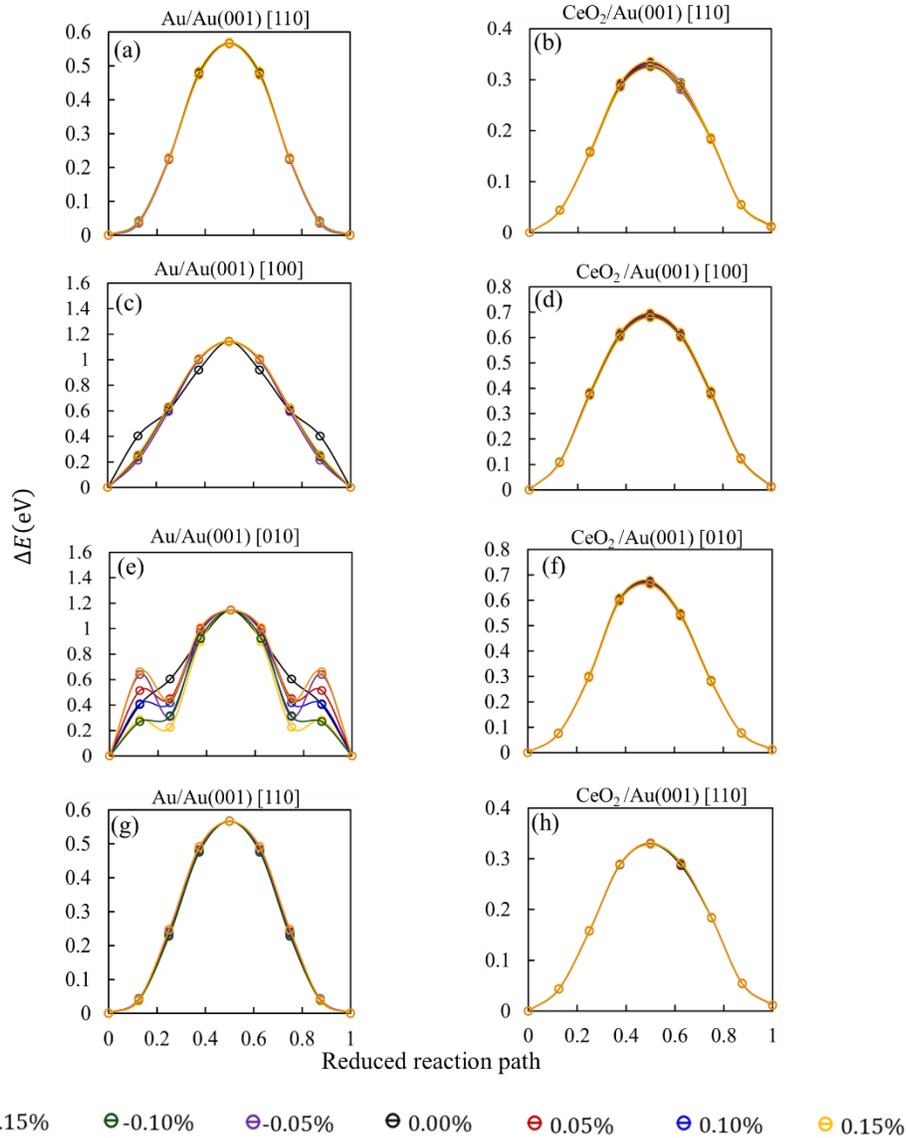

Fig. 7. Energy profile along the hopping direction of Au adatom (left panels) and CeO$_2$ admolecule (right panels) on Au (001) surface under strain. Panels (a,b), (c,d) and (e,f) show the effect of $\varepsilon_{11}$ applied in the [100] direction on the energy profiles for transitions along the [110], [100] and [010] directions, respectively. Panels (g,h) show the effect of shear strain on the energy profile for transition in the [110] direction. The color key under the figure refers to the value of the applied strain.

Fig. 8 shows the effects of uniaxial and shear strains on the activation barriers for hopping in various directions of Au adatom (upper row) and CeO$_2$ admolecule (lower row) on Au (001) surface. The uniaxial strain is applied in the [100] direction and the shear strain is in the plane of the surface (001). Overall, $\frac{\partial \Delta E^d}{\partial \varepsilon_{11}}$ and $\frac{\partial \Delta E^d}{\partial \varepsilon_{12}}$ are larger for CeO$_2$ admolecule than Au adatom, indicating a stronger sensitivity of the activation barrier of the CeO$_2$ admoleucle to surface strain. Turning attention to diffusion along [110] direction, although a CeO$_2$ admolecule generally exhibits lower activation barriers than an Au adatom, sufficiently large tensile strain on the order of, say, 4%-5%, can invert this trend, yielding higher activation energy barrier of CeO$_2$ admolecule than Au adatom. Note that the uniaxial strain $\varepsilon_{11}$ in Fig. 8 is applied along the [100] direction. As can be shown in the figure, the effect of strain on activation barriers in different directions varies, thus resulting in anisotropic diffusion. For example, considering Au adatom diffusion along [100] on (001) surface, the components, $\frac{\partial \Delta E^{[100]}}{\partial \varepsilon_{11}}$ and $\frac{\partial \Delta E^{[100]}}{\partial \varepsilon_{22}}$, will have 2.4714 eV and 1.642 eV, respectively. In contrast, the diffusion along [010] these values are interchanged yielding $\frac{\partial \Delta E^{[010]}}{\partial \varepsilon_{11}}$ and $\frac{\partial \Delta E^{[010]}}{\partial \varepsilon_{22}}$ being 1.642 eV and 2.4714 eV, respectively.

Fig. 8(b) and (d), respectively, show that positive shear increases the diffusion barrier along [110] direction while lowering its value along [$\bar{1}$10]. In theory, the magnitude of $\frac{\partial \Delta E^{[110]}}{\partial \varepsilon_{12}}$ and $\frac{\partial \Delta E^{[\bar{1}10]}}{\partial \varepsilon_{12}}$ are expected to be the same but with opposite sign. In Fig. 8(b) and (d), however, the magnitude of $\frac{\partial \Delta E^{[110]}}{\partial \varepsilon_{12}}$ and $\frac{\partial \Delta E^{[\bar{1}10]}}{\partial \varepsilon_{12}}$ are slightly different. (Note that $\gamma_{12} = 2\varepsilon_{12}$ and thus $\frac{\partial \Delta E^d}{\partial \varepsilon_{12}} = 2 \frac{\partial \Delta E^d}{\partial \gamma_{12}}$.)

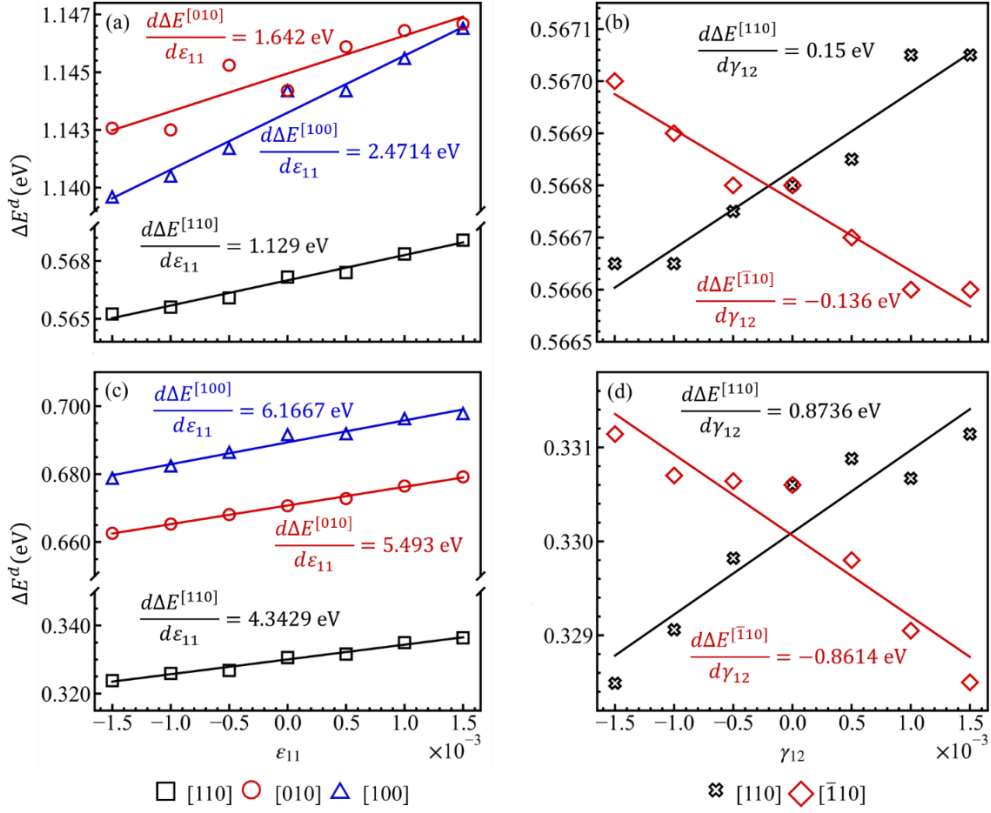

Fig. 8. Activation energies of Au adatom (upper row) and CeO$_2$ admolecule (lower row) hopping on Au (001) surface as a function of applied strain. Panels (a,c) show the effect of uniaxial strain applied in the [100] direction on activation energies for hopping in the [110], [100] and [010] directions, while panels (b,d) show the effect of shear strain in the (001) surface on the activation energies for hopping in the [110] and [$\bar{1}$10] directions. The symbols at the bottom of the figure refer to the hopping directions.

The energy profile along the minimum energy paths for hopping of Au atoms and CeO$_2$ molecules on Ce-terminated CeO$_2$ surface are shown in Fig. S1. (The reader is referred to the supplementary information for details.) In Fig. S2, the top and bottom rows show the effect of strain on activation energies for diffusion of Au and CeO$_2$ admolecule, respectively, on the Ce-terminated CeO$_2$ (001) surface. Unlike Au hopping of on Au (001), Au adatom hopping on CeO$_2$ surface exhibits an activation barrier that is higher under compression and lower under tension, which is shown in Fig. S2(a). Along the direction of applied strain, [100], increasing tensile strain reduces the activation barrier relative to diffusion along [010]. In contrast, as seen in Fig S2(c), the activation barrier for CeO$_2$ admolecule is lower in compression than in tension, which is similar to the case of hopping of the same molecule on Au (001). As shown in Fig. S2(b) and (d), the activation energy along [110] is lower than that along [$\bar{1}$10] when the shear strain is positive, and vice versa.

Additional results on energy profile along the minimum energy paths and diffusion barriers of Au and $CeO_2$ adsorbates on MgO (001) and both SrO and $TiO_2$ terminated surfaces of $SrTiO_3$ (001) are provided in supplementary information (see Fig. S3, S5 and S6). The effect of applied normal and shear strain on activation energies of Au and $CeO_2$ adsorbates on MgO (001) and both SrO and $TiO_2$ terminated surfaces of $SrTiO_3$ (001) are illustrated in Fig. S4, S7 and S8, respectively. Again, the reader is referred to the supplementary information for details. The impact of the elastic strain is shown to induce anisotropy of the diffusion barriers, which is expected to impact film growth.

Based on the linear fittings in Figs. 8, S2, S4, S7 and S8, the diffusion barrier is interpolated as function of strain in the form

$$\Delta E^d(\varepsilon_{11}, \varepsilon_{22}) = \Delta E^d(0) + \frac{\partial \Delta E^d}{\partial \varepsilon_{11}} \varepsilon_{11} + \frac{\partial \Delta E^d}{\partial \varepsilon_{11}} \varepsilon_{22} . \qquad (23)$$

To validate our model for the dependence of diffusion barrier on strain, see Eq. (23), we performed NEB calculations of $\Delta E^d$ on uniaxially- and biaxially- strained surfaces with strain values of 0.5% and -0.5%. The resulting activation energies for Au adatom and $CeO_2$ admolecule hopping along different directions are summarized in Tables S9-S12. The strain coefficient matrix $\frac{\partial \Delta E^d}{\partial \varepsilon_{ij}}$ is constructed from linear fits of $\Delta E^d$ versus strain as shown in Figs. 8, S2, S4, S7 and S8. For diffusion along [110], symmetry requires $\frac{\partial \Delta E^{[110]}}{\partial \varepsilon_{11}} = \frac{\partial \Delta E^{[110]}}{\partial \varepsilon_{22}}$ and $\frac{\partial \Delta E^{[\bar{1}10]}}{\partial \varepsilon_{11}} = \frac{\partial \Delta E^{[\bar{1}10]}}{\partial \varepsilon_{22}}$. As for the shear strain effect, the same requires $\frac{\partial \Delta E^{[\bar{1}10]}}{\partial \varepsilon_{12}} = -\frac{\partial \Delta E^{[110]}}{\partial \varepsilon_{12}}$. On the other hand, regarding the diffusion along [100] and [010] on Au, CeO$_2$ and MgO surfaces, no direct correlation between the applied shear and activation energies was observed. Therefore, it is assumed that $\frac{\partial \Delta E^{[100]}}{\partial \varepsilon_{12}} = \frac{\partial \Delta E^{[010]}}{\partial \varepsilon_{12}} = 0$. The same was found for hopping along the [110] direction on SrTiO$_3$ surface. In certain cases, such as Au adatom CeO$_2$ admolecule diffusion on biaxially strained surfaces like CeO$_2$ and SrTiO$_3$ surfaces, noticeable differences are observed between the interpolated values and the ones computed using NEB. These discrepancies likely arise due to neglecting the higher order terms in the energy expansion, i.e., terms of the form $\frac{\partial^2 \Delta E^d}{\partial \varepsilon_{11} \partial \varepsilon_{11}}$, $\frac{\partial^2 \Delta E^d}{\partial \varepsilon_{22} \partial \varepsilon_{22}}$ and $\frac{\partial^2 \Delta E^d}{\partial \varepsilon_{11} \partial \varepsilon_{22}}$. As mentioned earlier, these refinements are beyond the scope of the current work.

In the limit of small deformation, our results show a clear dependency of surface, binding, and diffusion activation energies on the elastic strain. During film growth by self-assembly methods such as PLD or Molecular Beam Epitaxy (MBE), these findings highlight the potential role of strain engineering in driving self-organization and/or phase separation. The simple linear interpolation for adatom/admolecule binding and diffusion energies based on Eq. (20), see also Eqs. (22) and (23), can be integrated into mesoscale models of diffusion, nucleation and growth, by evaluating the binding and diffusion barriers as functions of the local strain fields arising from lattice mismatch, which may vary from point to point on the surface [47,85]. This study thus highlights a potential role of elastic strain in tailoring the kinetic rates and binding energies to modulate island nucleation and film growth. In addition, this work allows us to investigate the competition between short-range chemical interaction and long-range elastic forces during the growth of thin films [21,87], thus allowing for multiphase thin film design.

## 5. Concluding remarks

First-principles calculations were performed to investigate the impact of surface strain on binding and activation energies of Au adatom and $CeO_2$ admolecule on Au and various oxide surfaces. Specifically, the Au (001), Ce-terminated $CeO_2$ (001), MgO (001), and both SrO and $TiO_2$-terminated $SrTiO_3$ (001) surfaces were considered. Binding energies were evaluated at the minimum energy adsorption sites, taken here to be the hollow sites. The energy landscape along the minimum energy diffusion paths were examined for the [110], [$\bar{1}$10], [100] and [010] directions on all surfaces considered in this work, and corresponding activation barriers were calculated using the NEB method.

This current work showed that binding and activation energies depend on the applied strain, and as expected, on the material and surface type, and the type of adsorbed species. For instance, Au adatom was found to have higher (lower) binding energy on Au surface than $CeO_2$ Ce-terminated surface when tensile (compression) strain is applied along [100]. It was also found that Au adatom hopping in any direction has higher (lower) activation energy on Au surface compared with Ce-terminated $CeO_2$ surface when tensile (compression) strain is applied along [100]. The current work further reveals that strain application to surfaces can result in anisotropy of diffusion barriers, indicating the possible major role of strain in surface diffusion, island nucleation, and film growth. In addition, the results show that the energetics of binding and diffusion depend linearly on surface strain up to fairly large values (of around 0.5%) in most cases and that, in some cases, refinements are required by incorporating higher order coefficients for strain dependence of the energy quantities of interest. Finally, in performing the calculations for the binding and hopping of $CeO_2$ molecules, these molecules were considered rigid. This assumption can be relaxed in future investigations.

With regard to strain effects on the binding and activation energies of adsorbates, the current work clearly asserts the possibility to use strain as a control apparatus in island nucleation and patterning during thin film growth of single and multiple phases. Indeed, the current authors have used the current results in the simulation of multi-phase film growth and demonstrated a significant effect of strain [88].

**Supplementary material**

The supplementary material contains additional results and discussion on the activation energies computed using the NEB method for Au adatoms and $CeO_2$ admolecules on Ce-terminated $CeO_2$, MgO, and SrO- and $TiO_2$-terminated $SrTiO_3$ surfaces.

**Acknowledgements**

This work is supported by the U.S. Department of Energy, Office of Science, Basic Energy Sciences (BES) under award DE-SC0020077 at Purdue University. We would like to thank Kyle Starkey for reviewing and giving feedback on an early version of this manuscript.

**Conflict of interest**

The authors have no conflicts to disclose.

# Strain effects on the binding and diffusion energies of Au adatoms and CeO$_2$ admolcules on Au, CeO$_2$, MgO and SrTiO$_3$ surfaces

Ahmad Ahmad, Ying-Cheng Chen, Jie Peng[‡], Anter El-Azab*

School of Materials Engineering, Purdue University, West Lafayette, IN 47907, USA

[‡]Was with Purdue University during this study

*Corresponding author: Anter El-Azab, aelazab@purdue.edu

The surface energies obtained directly from DFT calculations and those extrapolated using Eq. (21) are summarized in Tables S1–S4. Good agreement between the extrapolated values and those computed by DFT is observed.

Table S1. Surface energies at 0.5% uniaxial strain.

| Surface | (DFT) $E^s$ (J/m$^2$) | (Interpolated) $E^s$ (J/m$^2$) |
|---|---|---|
| Au (001) | 0.8822 | 0.8826 |
| CeO$_2$ (001) | 4.1576 | 4.1592 |
| MgO (001) | 0.9346 | 0.9352 |
| SrO (001) | 1.0809 | 1.0798 |
| TiO$_2$ (001) | 1.0846 | 1.0856 |

Table S2. Surface energies at -0.5% uniaxial strain.

| Surface | (DFT) $E^s$ (J/m$^2$) | (Interpolated) $E^s$ (J/m$^2$) |
|---|---|---|
| Au (001) | 0.8707 | 0.8714 |
| CeO$_2$ (001) | 4.2023 | 4.2028 |
| MgO (001) | 0.9100 | 0.9108 |
| SrO (001) | 1.0686 | 1.0662 |
| TiO$_2$ (001) | 1.0679 | 1.0684 |

Table S3. Surface energies at 0.5% biaxial strain.

| Surface | (DFT) $E^s$ (J/m$^2$) | (Interpolated) $E^s$ (J/m$^2$) |
|---|---|---|
| Au (001) | 0.8862 | 0.8881 |
| CeO$_2$ (001) | 4.1338 | 4.1374 |
| MgO (001) | 0.9467 | 0.9474 |
| SrO (001) | 1.0862 | 1.0867 |
| TiO$_2$ (001) | 1.0922 | 1.0942 |

Table S4. Surface energies at -0.5% biaxial strain.

| Surface | (DFT) $E^s$ (J/m$^2$) | (Interpolated) $E^s$ (J/m$^2$) |
|---|---|---|
| Au (001) | 0.8632 | 0.8659 |
| CeO$_2$ (001) | 4.2231 | 4.2246 |
| MgO (001) | 0.8975 | 0.8986 |
| SrO (001) | 1.0617 | 1.0593 |
| TiO$_2$ (001) | 1.0589 | 1.0599 |

The binding energies at large strain (±0.5%) calculated directly from DFT are compared to those extrapolated using Eq. (2*2*) in Tables S5–S8.

Table S5. Binding energies of Au and CeO$_2$ at the minimum energy adsorption site at 0.5% uniaxial strain.

| Surface | Adsorbed specie | (DFT) $E^b$ (eV) | (Extrapolated) $E^b$ (eV) |
|---|---|---|---|
| Au (001) | Au | -2.9950 | -2.9947 |
|  | CeO$_2$ | -1.2797 | -1.2789 |
| CeO$_2$ (001) | Au | -4.1780 | -4.1784 |
|  | CeO$_2$ | -4.5035 | -4.5025 |
| MgO (001) | Au | -1.0108 | -1.0093 |
|  | CeO$_2$ | -3.4261 | -3.4287 |
| SrO (001) | Au | -1.3654 | -1.3650 |
|  | CeO$_2$ | -4.4421 | -4.4527 |
| TiO$_2$ (001) | Au | -0.4648 | -0.4692 |
|  | CeO$_2$ | -6.3933 | -6.3908 |

Table S6. Binding energies of Au and CeO$_2$ at the minimum energy adsorption site at -0.5% uniaxial strain.

| Surface | Adsorbed specie | (DFT) $E^b$ (eV) | (Extrapolated) $E^b$ (eV) |
|---|---|---|---|
| Au (001) | Au | -2.9691 | -2.9673 |
|  | CeO$_2$ | -1.2564 | -1.2551 |
| CeO$_2$ (001) | Au | -4.1852 | -4.1876 |
|  | CeO$_2$ | -4.5057 | -4.5055 |
| MgO (001) | Au | -0.9977 | -1.0007 |
|  | CeO$_2$ | -3.4029 | -3.4113 |
| SrO (001) | Au | -1.3727 | -1.3730 |
|  | CeO$_2$ | -4.4195 | -4.4094 |
| TiO$_2$ (001) | Au | -0.4555 | -0.4549 |
|  | CeO$_2$ | -6.3749 | -6.3752 |

Table S7. Binding energies of Au and CeO$_2$ at the minimum energy adsorption site at 0.5% biaxial strain.

| surface | Adsorbed specie | (DFT) $E^b$ (eV) | (Extrapolated) $E^b$ (eV) |
|---|---|---|---|
| Au (001) | Au | -3.0087 | -3.0085 |
|  | CeO$_2$ | -1.2910 | -1.2908 |
| CeO$_2$ (001) | Au | -4.1739 | -4.1737 |
|  | CeO$_2$ | -4.5008 | -4.5010 |
| MgO (001) | Au | -1.0178 | -1.0136 |
|  | CeO$_2$ | -3.4455 | -3.4375 |
| SrO (001) | Au | -1.3619 | -1.3611 |
|  | CeO$_2$ | -4.4482 | -4.4743 |
| TiO$_2$ (001) | Au | -0.4702 | -0.4763 |
|  | CeO$_2$ | -6.4408 | -6.3986 |

Table S8. Binding energies of Au and CeO$_2$ at the minimum energy adsorption site at -0.5% biaxial strain.

| Surface | Adsorbed specie | (DFT) $E^b$ (eV) | (Extrapolated) $E^b$ (eV) |
|---|---|---|---|
| Au (001) | Au | -2.9568 | -2.9535 |
|  | CeO$_2$ | -1.2439 | -1.2432 |
| CeO$_2$ (001) | Au | -4.1911 | -4.1923 |
|  | CeO$_2$ | -4.5075 | -4.5070 |
| MgO (001) | Au | -0.9927 | -0.9964 |
|  | CeO$_2$ | -3.3877 | -3.4025 |
| SrO (001) | Au | -1.3764 | -1.3770 |
|  | CeO$_2$ | -4.4138 | -4.3877 |
| TiO$_2$ (001) | Au | -0.4513 | -0.4477 |
|  | CeO$_2$ | -6.3278 | -6.3674 |

The energy profiles along the minimum energy paths for diffusion of Au and CeO$_2$ on Ce-terminated CeO$_2$ surface are shown in Fig. S1. In parts (d) and (f) of the figure, local minima at reaction coordinate value of 0.5 were observed. In computing the activation barriers, however, these local minima are ignored, and the peak values of the curves are considered in barrier calculations. In the case of CeO$_2$ diffusion on the Ce-terminated CeO$_2$ surface, as shown in Fig. S1(h), it is noticed that the shear has the largest impact on the minimum energy path.

In Fig. S2, diffusion barriers of both Au adatom and CeO$_2$ admolecule diffusion along [110], [100], [010] and [$\bar{1}$10] on CeO$_2$ (001) surface are presented. The activation energy barriers of Au adatom hopping in the [100] and [010] directions on CeO$_2$ (001) are in proximity to the values obtained on Au (001) surface. Moreover, as seen on the Au (001) surface, in general, CeO$_2$ the diffusion barrier is more sensitive to both uniaxial or shear strain on the Ce-terminated CeO$_2$ surface than Au adatom diffusion. Based on linear interpolation, the most pronounced effect is observed under shear strain, where ±1.0% strain can modify the activation energy barrier of CeO$_2$ admolecule diffusion by 0.2 eV, see Fig. S2(d).

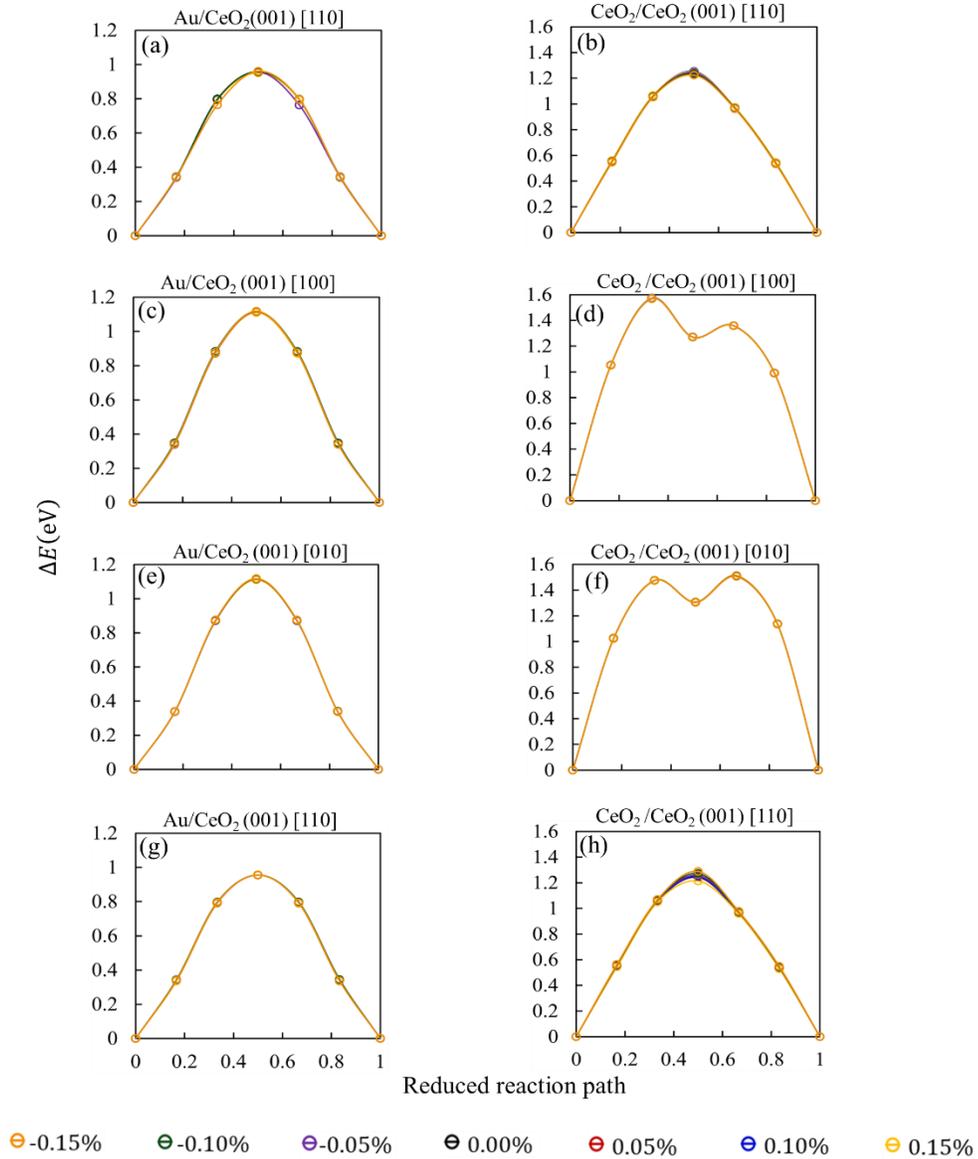

Fig. S1. Energy profiles along the hopping direction of Au adatom (left panels) and $CeO_2$ admolecule (right panels) on Ce-terminated $CeO_2$ (001) surface under strain. Panels (a,b), (c,d) and (e,f) show the effect of uniaxial strain $\varepsilon_{11}$ applied in the [100] direction on the energy profiles for transitions along [110], [100] and [010] directions, respectively. Panels (g,h) show the effect of shear strain on the energy profile for transition in the [110] direction. The color key under the figure refers to the value of the applied strain.

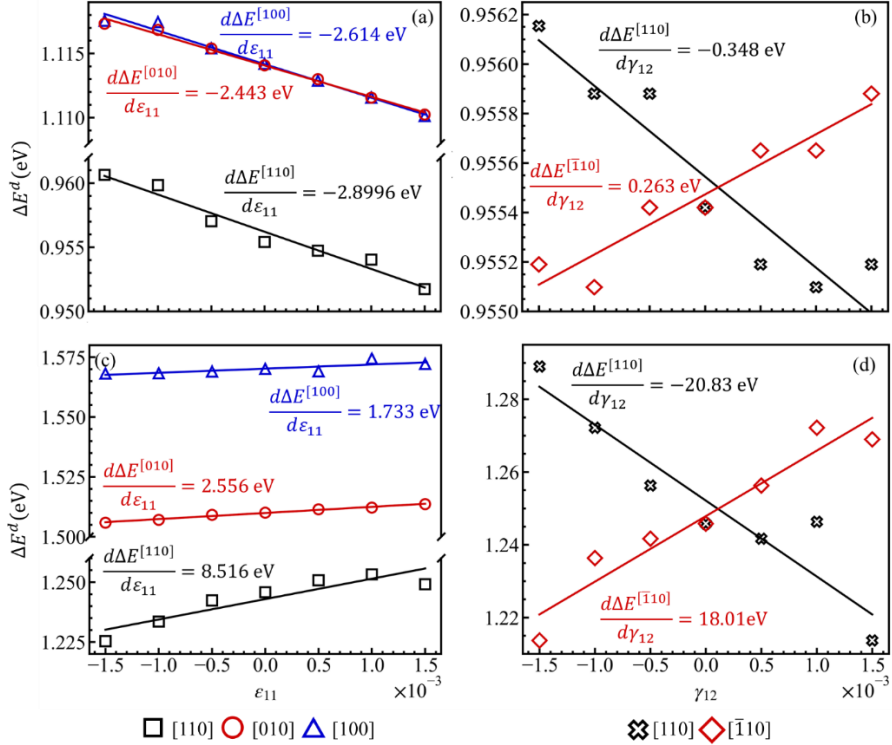

Fig. S2. Activation energies of Au adatom (upper row) and CeO$_2$ admolecule (lower row) hopping on Ce-terminated CeO$_2$ (001) surface as a function of applied strain. Panels (a,c) show the effect of uniaxial strain applied in the [100] direction on activation energies for hopping in the [110], [100] and [010] directions, while panels (b,d) show the effect of shear strain in the (001) surface on the activation energies for hopping in the [110] and [$\bar{1}$10] directions. The symbols at the bottom of the figure refer to the hopping directions.

As reported previously [1,2], MgO has been used as the substrate for growing film systems. For this reason, we have also examined the impact of strain on the diffusion of Au adatom and CeO$_2$ admolecule on MgO (001). The effect of strain on the energy profiles of hopping adatoms and molecules on this surface is shown in Fig. S3.

Fig. S4(a,b) and (c,d) shows the effect of strained MgO surface on the activation energies of Au adatom and CeO$_2$ admolecule hopping diffusion mechanism, respectively. For Au diffusion along [110], tensile strain lowers the activation barrier compared to compression as seen in Fig. S4(a). In contrast, CeO$_2$ diffusion along [110] exhibits the opposite trend, with compression reducing the barrier relative to tension as shown in Fig. S4(c). Along [100] and [010], both Au and CeO₂ diffusions display higher activation barriers under tensile strain than compression. The effect of shear strain on MgO, as shown in Fig. S4(b) and (d), follows trends similar to those observed on the Au surface.

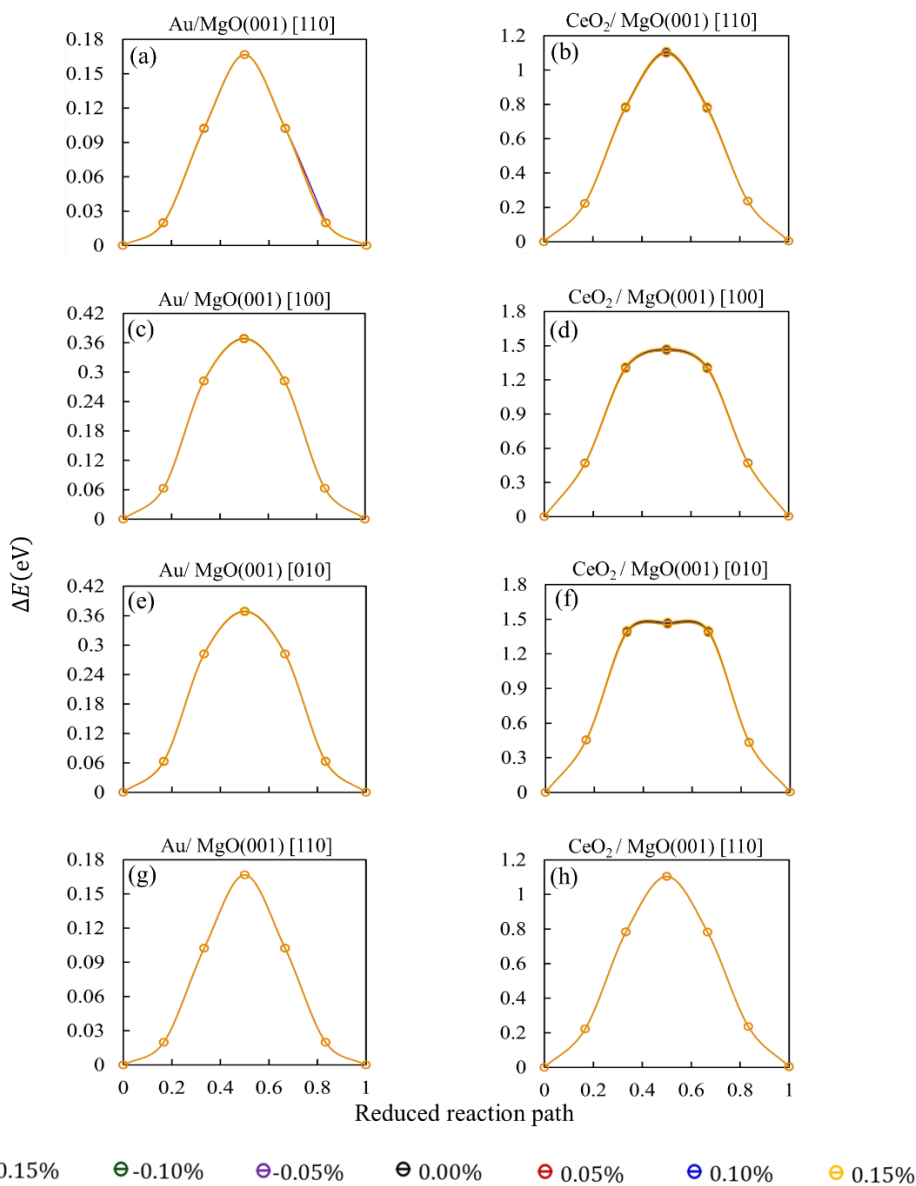

Fig. S3. Energy profiles along the hopping direction of Au adatom (left panels) and $CeO_2$ admolecule (right panels) on MgO (001) surface under strain. Panels (a,b), (c,d) and (e,f) show the effect of uniaxial strain $\varepsilon_{11}$ applied in the [100] direction on the energy profiles for transitions along [110], [100] and [010] directions, respectively. Panels (g,h) show the effect of shear strain on the energy profile for transition in the [110] direction. The color key under the figure refers to the value of the applied strain.

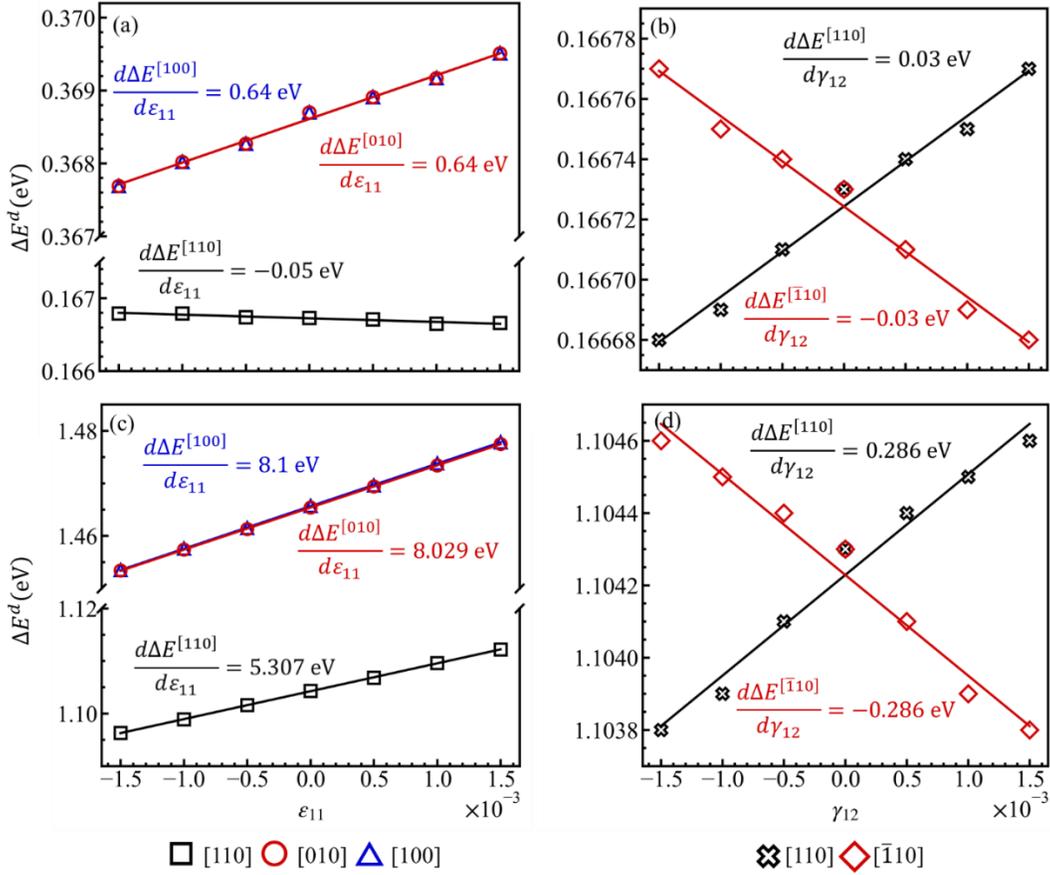

Fig. S4. Activation energies of Au adatom (upper row) and CeO$_2$ admolecule (lower row) hopping on MgO (001) surface as a function of applied strain. Panels (a,c) show the effect of uniaxial strain applied in the [100] direction on activation energies for hopping in the [110], [100] and [010] directions, while panels (b,d) show the effect of shear strain in the (001) surface on the activation energies for hopping in the [110] and [$\bar{1}$10] directions. The symbols at the bottom of the figure refer to the hopping directions.

Figs. S5 and S6 present the energy profiles along the hopping directions on SrO- and TiO$_2$-terminated SrTiO$_3$ (001) surfaces, respectively. For the SrO-terminated surface, a noticeable dependence on strain is observed along [110] as seen in Fig. S5(a), where decreasing strain narrows the energy profile. Examining Fig. S5(b–j), for other diffusion paths, the variations are limited to the third decimal place, rendering them indistinguishable at the displayed scale. As seen in Fig. S6, for the TiO$_2$-terminated surface, Au adatom diffusion shows pronounced strain effects. Fig. S6(a) and (h), respectively, show that, along [110] direction both tensile and negative shear strains increase the activation barrier. Significant strain dependence is also observed for diffusion along the [100] and [010] directions under normal strain as seen in Fig. S6(b) and (d) as well as shear strain as seen in Fig. S6(f), with the shear strain equally affecting energies for the [100] and [010] directions. In contrast, $CeO_2$ molecule hopping shows a weaker sensitivity to applied strain, with only minor changes observed in Figs. S6(c), (e), and (g).

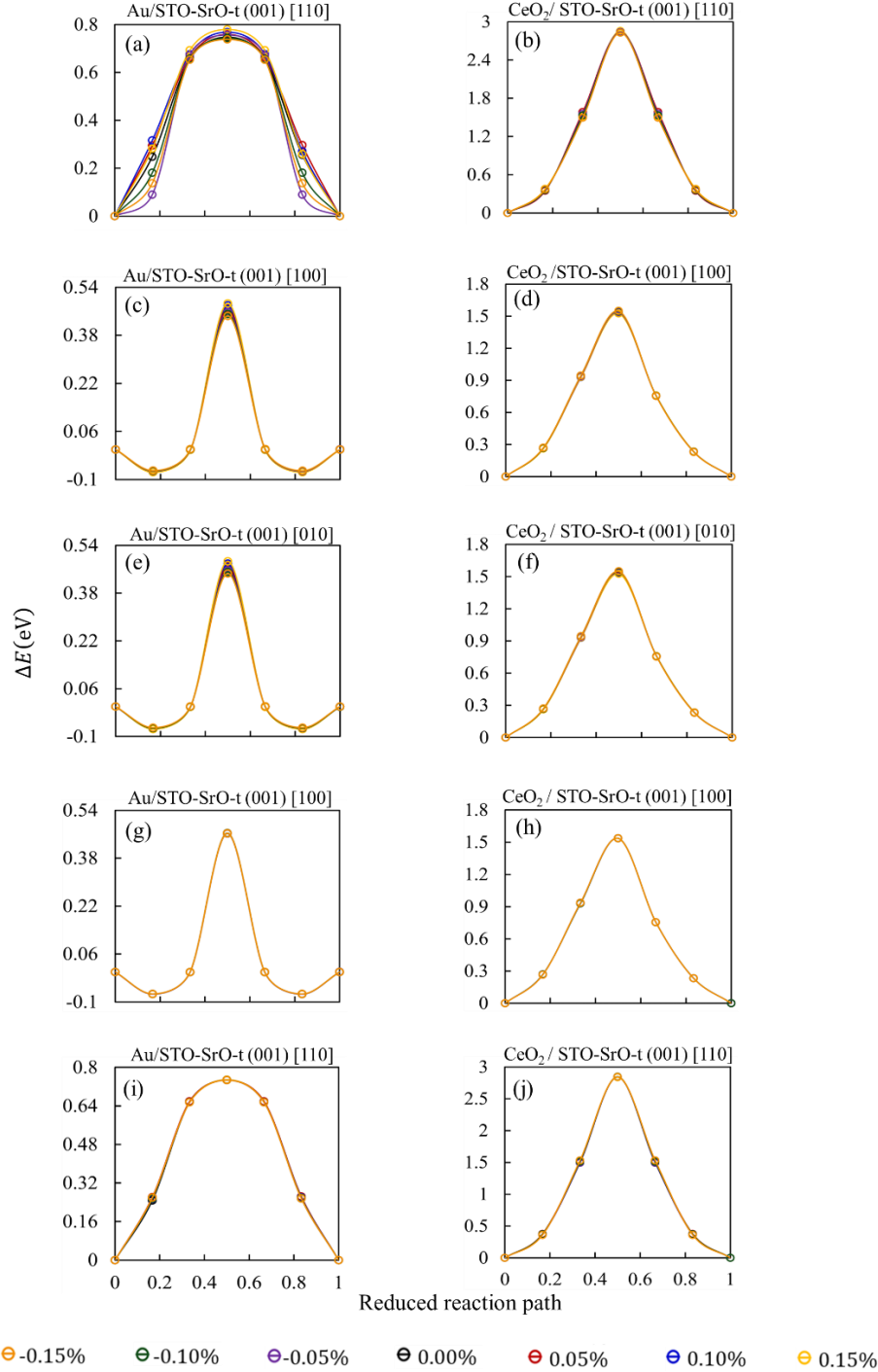

Fig. S5. Energy profile along the hopping direction of Au adatom (left panels) and $CeO_2$ admolecule (right panels) on SrO-terminated $SrTiO_3$ (001) free surface under strain. Panels (a,b), (c,d) and (e,f) show the effect of uniaxial strain $\varepsilon_{11}$ applied in the [100] direction on the energy profiles for transitions along [110], [100] and [010], respectively. Panels (g,h) and (i,j) show the effect of shear strain on the energy profile for transition in [100] and [110] directions, respectively. The color key under the figure refers to the value of the applied strain.

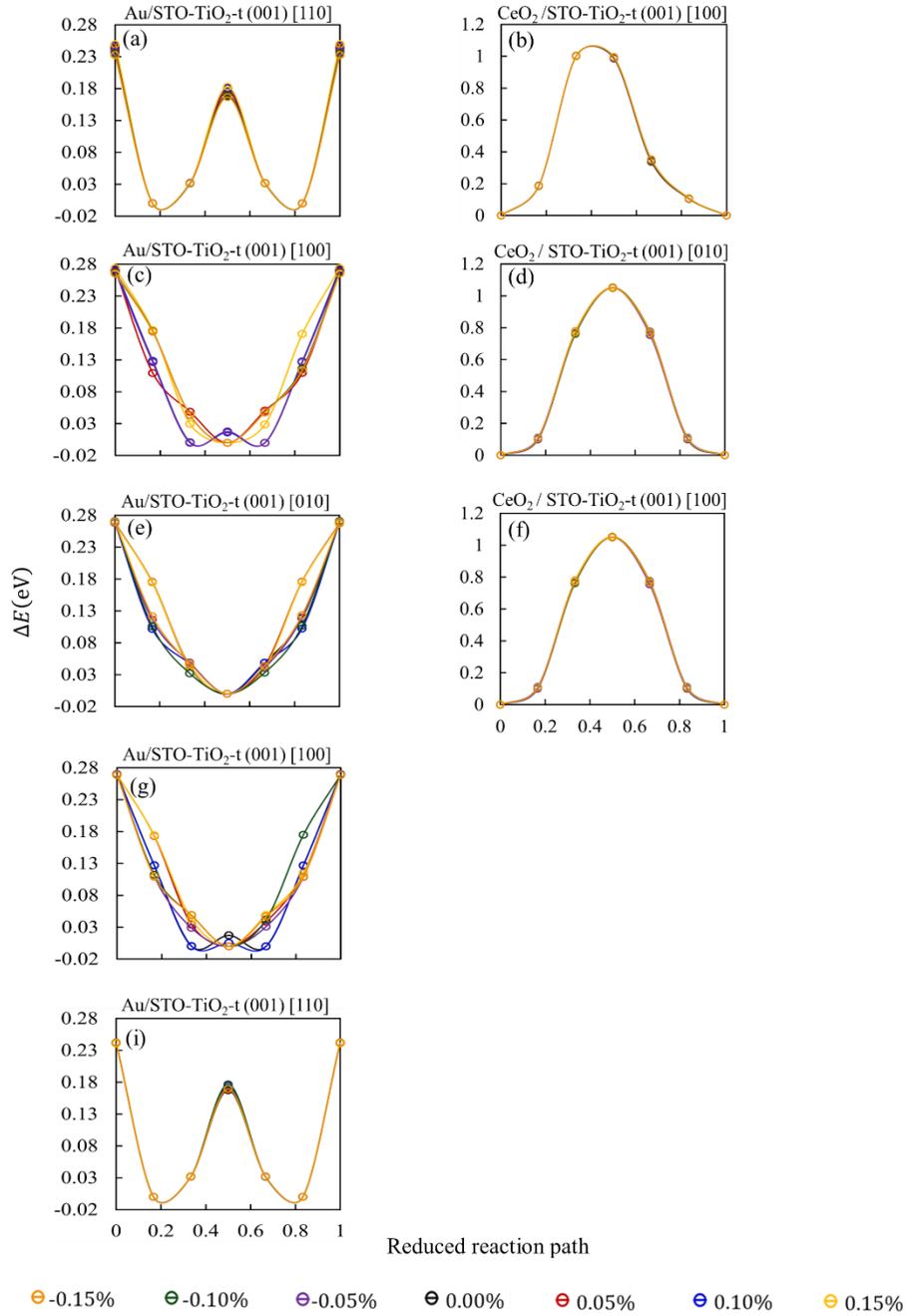

Fig. S6. Energy profiles along the hopping direction of Au adatom (left panels) and $CeO_2$ admolecule (right panels) on $TiO_2$-terminated $SrTiO_3$ (001) free surface under strain. Panels (a), (b,c) and (d,e) show the effect of uniaxial strain $\varepsilon_{11}$ applied in the [100] direction on the energy profiles for transitions along [110], [100] and [010], respectively. Panels (f,g) and (h) show the effect of shear strain on the energy profile for transition in [100] and [110] directions, respectively. The color key under the figure refers to the value of the applied strain.

Figs. S7 and S8 present the effect of strain on Au adatom and $CeO_2$ admolecule diffusion on SrO- and $TiO_2$-terminated $SrTiO_3$ (001) surfaces, respectively. For Au on the SrO-terminated surface as seen in Fig. S7(a), tensile strain increases the activation barrier for diffusion along the [110], [100], and [010] directions. As shown in Fig. S7(b), under shear strain, diffusion along [100] and [010] shows equivalent behavior, with positive strain lowering the barrier, while diffusion along [110] is similarly affected. For CeO₂ diffusion as shown in Fig. S7(c), tensile strain raises the activation barrier along [110] but lowers it along [100] and [010]. Under shear as seen in Fig. S7(d), CeO₂ diffusion along [100] and [110] mirrors the trend observed for Au adatoms.

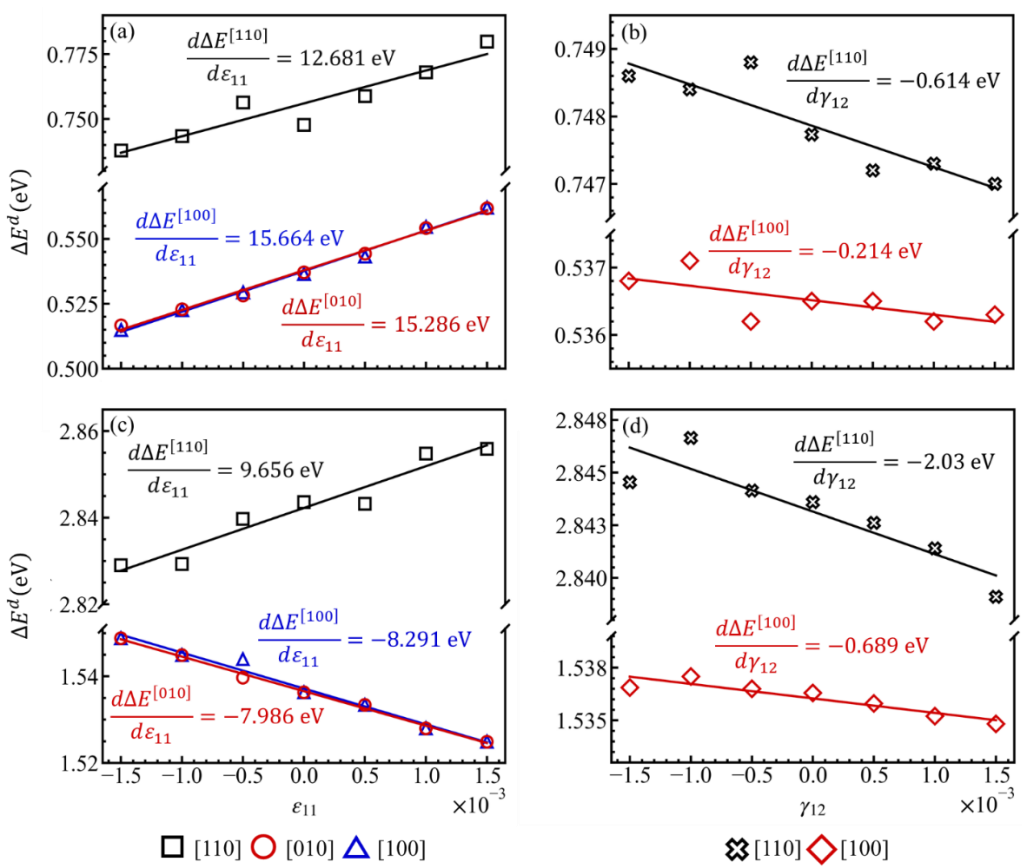

Fig. S7. Activation energies of Au adatom (upper row) and CeO₂ admolecule (lower row) hopping on SrO-terminated SrTiO₃ (001) surface as a function of applied strain. Panels (a,c) show the effect of uniaxial strain applied in the [100] direction on activation energies for hopping in the [110], [100] and [010] directions, while panels (b,d) show the effect of shear strain in the (001) surface on the activation energies for hopping in the [110] and [100] directions. The symbols at the bottom of the figure refer to the hopping directions.

The minimum energy paths for Au adatom and CeO₂ admolecule diffusion on the TiO₂-terminated SrTiO₃ (001) surface are shown in Fig. S8 left and right columns, respectively.

Examining Fig. S8(a), for Au diffusion along [110], tensile strain increases the activation barrier, whereas along [100] the barrier decreases with tensile strain, in contrast to [010], where it increases. Under positive shear, as seen in Fig. S8(b), diffusion along [100] exhibits a higher barrier, while along [110] the barrier is reduced. As observed in Fig. S8(c) and (d), for $CeO_2$ diffusion, both tensile and positive shear strains increase the activation barrier along [100] and [010].

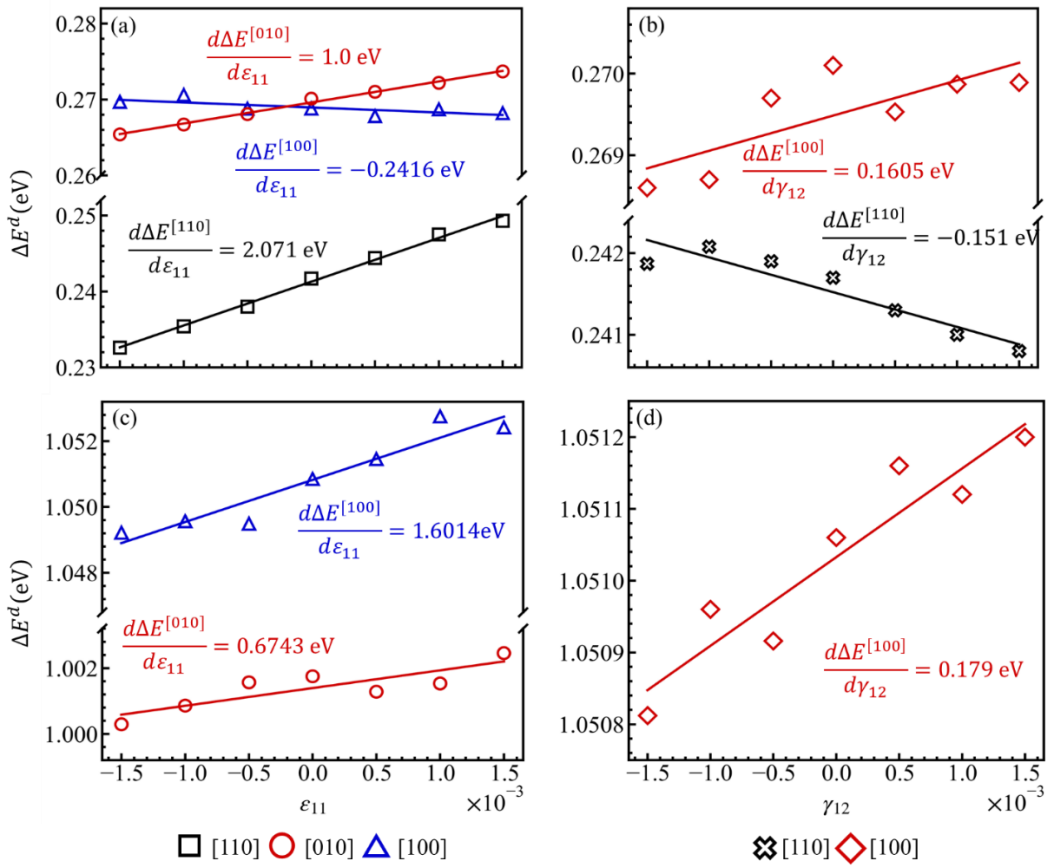

Fig. S8. Activation energies of Au adatom (upper row) and $CeO_2$ admolecule (lower row) hopping on $TiO_2$-terminated $SrTiO_3$ (001) surface as a function of applied strain. Panels (a,c) show the effect of uniaxial strain applied in the [100] direction on activation energies for hopping in the [110], [100] and [010] directions, while panels (b,d) show the effect of shear strain in the (001) surface on the activation energies for hopping in the [110] and [100] directions. The symbols at the bottom of the figure refer to the hopping directions.

The impact of large strain on the activation energy barriers for different hopping directions is evaluated and results obtained from direct NEB calculations and interpolation based on Eq. (23) are compared. Tables S9 and S10 show the impact of applied uniaxial strain at +0.5% and -0.5%, respectively, on the activation energies of Au adatom and CeO2 admolecule hopping different surfaces along a given direction. Tables S11 and S12 show the impact of applied biaxial strain at +0.5% and -0.5%, respectively, on the activation energies of Au adatom and CeO2 admolecule hopping different surfaces along a given direction. As can be seen, the interpolated activation energies at higher applied strain are in good agreement with the values directly computed from NEB.

Table S9. Activation energy barrier for Au adatom and CeO2 admolecule hopping on different surfaces under a uniaxial normal strain of 0.5%.

| Surface | Adatom | Diffusion direction | (DFT) $\Delta E^d$ (eV) | (Interpolated) $\Delta E^d$ (eV) | Admolecule | Diffusion direction | (DFT) $\Delta E^d$ (eV) | (Interpolated) $\Delta E^d$ (eV) |
|---|---|---|---|---|---|---|---|---|
| Au (001) | Au | [110] | 0.5723 | 0.5724 | $CeO_2$ | [110] | 0.3568 | 0.3523 |
| | | [100] | 1.1512 | 1.1546 | | [100] | 0.7164 | 0.7008 |
| | | [010] | 1.1510 | 1.1504 | | [010] | 0.7007 | 0.6975 |
| $CeO_2$ (001) | Au | [110] | 0.9446 | 0.9409 | $CeO_2$ | [110] | 1.2629 | 1.2884 |
| | | [100] | 1.1042 | 1.1010 | | [100] | 1.6359 | 1.5788 |
| | | [010] | 1.1107 | 1.1020 | | [010] | 1.6286 | 1.5229 |
| MgO (001) | Au | [110] | 0.1666 | 0.1665 | $CeO_2$ | [110] | 1.1278 | 1.1308 |
| | | [100] | 0.3717 | 0.3719 | | [100] | 1.5028 | 1.5062 |
| | | [010] | 0.3717 | 0.3719 | | [010] | 1.5024 | 1.5055 |
| SrO (001) | Au | [110] | 0.8119 | 0.8211 | $CeO_2$ | [110] | 2.8709 | 2.8919 |
| | | [100] | 0.6070 | 0.6148 | | [100] | 1.5031 | 1.4948 |
| | | [010] | 0.6061 | 0.6135 | | [010] | 1.5033 | 1.4964 |
| $TiO_2$ (001) | Au | [110] | 0.1757 | 0.1779 | $CeO_2$ | [100] | 0.9973 | 1.0102 |
| | | [100] | 0.2728 | 0.2676 | | [010] | 1.0554 | 1.0545 |
| | | [010] | 0.2977 | 0.2751 | | | | |

Table S10. Activation energy barrier for Au adatom and CeO2 admolecule hopping on different surfaces under a uniaxial normal strain of -0.5%.

| Surface | Adatom | Diffusion direction | (DFT) $\Delta E^d$ (eV) | (Interpolated) $\Delta E^d$ (eV) | Admolecule | Diffusion direction | (DFT) $\Delta E^d$ (eV) | (Interpolated) $\Delta E^d$ (eV) |
|---|---|---|---|---|---|---|---|---|
| Au (001) | Au | [110] | 0.5617 | 0.5612 | $CeO_2$ | [110] | 0.3116 | 0.3089 |
| | | [100] | 1.1270 | 1.1298 | | [100] | 0.6652 | 0.6392 |
| | | [010] | 1.1276 | 1.1340 | | [010] | 0.6461 | 0.6425 |
| $CeO_2$ (001) | Au | [110] | 0.9752 | 0.9699 | $CeO_2$ | [110] | 1.1940 | 1.2032 |
| | | [100] | 1.1352 | 1.1272 | | [100] | 1.4887 | 1.5614 |
| | | [010] | 1.1318 | 1.1264 | | [010] | 1.4535 | 1.4973 |
| MgO (001) | Au | [110] | 0.1668 | 0.1670 | $CeO_2$ | [110] | 1.0741 | 1.0778 |
| | | [100] | 0.3655 | 0.3655 | | [100] | 1.4224 | 1.4252 |
| | | [010] | 0.3655 | 0.3655 | | [010] | 1.4222 | 1.4253 |
| SrO (001) | Au | [110] | 0.6752 | 0.6943 | $CeO_2$ | [110] | 2.8084 | 2.7953 |
| | | [100] | 0.4657 | 0.4582 | | [100] | 1.5691 | 1.5778 |
| | | [010] | 0.4619 | 0.4607 | | [010] | 1.5692 | 1.5762 |
| $TiO_2$ (001) | Au | [110] | 0.1340 | 0.1571 | $CeO_2$ | [100] | 0.9875 | 0.9942 |
| | | [100] | 0.2760 | 0.2651 | | [010] | 1.0267 | 1.0477 |
| | | [010] | 0.2607 | 0.2700 | | | | |

Table S11. Activation energy barrier for Au adatom and CeO2 admolecule hopping on different surfaces under applied biaxial strain of 0.5%.

| Surface | Adatom | Diffusion direction | (Interpolated) $\Delta E^d$ (eV) | (DFT) $\Delta E^d$ (eV) | Admolecule | Diffusion direction | (Interpolated) $\Delta E^d$ (eV) | (DFT) $\Delta E^d$ (eV) |
|---|---|---|---|---|---|---|---|---|
| Au (001) | Au | [110] | 0.5759 | 0.5781 | CeO2 | [110] | 0.3726 | 0.3740 |
|  |  | [100] | 1.1616 | 1.1628 |  | [100] | 0.7412 | 0.7283 |
|  |  | [010] | 1.1612 | 1.1628 |  | [010] | 0.7256 | 0.7499 |
| CeO2 (001) | Au | [110] | 0.9299 | 0.9264 | CeO2 | [110] | 1.3005 | 1.3310 |
|  |  | [100] | 1.0927 | 1.0888 |  | [100] | 1.6805 | 1.5915 |
|  |  | [010] | 1.0890 | 1.0889 |  | [010] | 1.6469 | 1.5315 |
| MgO (001) | Au | [110] | 0.1665 | 0.1662 | CeO2 | [110] | 1.1459 | 1.1574 |
|  |  | [100] | 0.3748 | 0.3751 |  | [100] | 1.5282 | 1.5463 |
|  |  | [010] | 0.3748 | 0.3751 |  | [010] | 1.5280 | 1.5460 |
| SrO (001) | Au | [110] | 0.8723 | 0.8845 | CeO2 | [110] | 2.9353 | 2.9402 |
|  |  | [100] | 0.6778 | 0.6913 |  | [100] | 1.5332 | 1.4549 |
|  |  | [010] |  | 0.6919 |  | [010] | 1.5333 | 1.4549 |
| TiO2 (001) | Au | [110] | 0.2430 | 0.1882 | CeO2 | [100] | 1.0098 | 1.0136 |
|  |  | [100] | 0.2803 | 0.2726 |  | [010] | 1.0687 | 1.0625 |
|  |  | [010] | 0.3505 | 0.2739 |  |  |  |  |

Table S12. Activation energy barrier for Au adatom and CeO2 admolecule hopping on different surfaces under applied biaxial strain of -0.5%.

| Surface | Adatom | Diffusion direction | (Interpolated) $\Delta E^d$ (eV) | (DFT) $\Delta E^d$ (eV) | Admolecule | Diffusion direction | (Interpolated) $\Delta E^d$ (eV) | (DFT) $\Delta E^d$ (eV) |
|---|---|---|---|---|---|---|---|---|
| Au (001) | Au | [110] | 0.5556 | 0.5555 | CeO2 | [110] | 0.3067 | 0.2872 |
|  |  | [100] | 1.1153 | 1.1216 |  | [100] | 0.6535 | 0.6117 |
|  |  | [010] | 1.1152 | 1.1216 |  | [010] | 0.6322 | 0.6333 |
| CeO2 (001) | Au | [110] | 0.9805 | 0.9844 | CeO2 | [110] | 1.1889 | 1.1606 |
|  |  | [100] | 1.1422 | 1.1394 |  | [100] | 1.4424 | 1.5487 |
|  |  | [010] | 1.1516 | 1.1395 |  | [010] | 1.4062 | 1.4887 |
| MgO (001) | Au | [110] | 0.1666 | 0.1672 | CeO2 | [110] | 1.0576 | 1.0512 |
|  |  | [100] | 0.3621 | 0.3623 |  | [100] | 1.3990 | 1.3851 |
|  |  | [010] | 0.3621 | 0.3623 |  | [010] | 1.3988 | 1.3848 |
| SrO (001) | Au | [110] | 0.597 | 0.6309 | CeO2 | [110] | 2.7802 | 2.7470 |
|  |  | [100] | 0.3960 | 0.3818 |  | [100] | 1.5381 | 1.6177 |
|  |  | [010] | 0.3964 | 0.3824 |  | [010] | 1.5383 | 1.6177 |
| TiO2 (001) | Au | [110] | 0.1221 | 0.1468 | CeO2 | [100] | 0.9759 | 0.9908 |
|  |  | [100] | 0.2756 | 0.2650 |  | [010] | 1.0119 | 1.0397 |
|  |  | [010] | 0.2770 | 0.2663 |  |  |  |  |